\RequirePackage{fix-cm}
\documentclass[smallextended]{svjour3}       
\smartqed  
\usepackage{graphicx}
\usepackage{amssymb,amsmath,amsfonts,latexsym}
\usepackage{mathtools}
\usepackage{url}
\usepackage[plainpages=false, bookmarks, bookmarksnumbered,
colorlinks, linktocpage=true, linkcolor=blue, citecolor=blue, filecolor=black,
urlcolor=blue]{hyperref}
\usepackage{natbib}
\setcitestyle{aysep={}} 
\newenvironment{sysmatrix}[1]
{\left(\begin{array}{@{}#1@{}}}
	{\end{array}\right)}
\newcommand{\ro}[1]{%
	\xrightarrow{\mathmakebox[\rowidth]{#1}}%
}
\newlength{\rowidth}
\AtBeginDocument{\setlength{\rowidth}{3em}}
\usepackage{indentfirst}
\usepackage{bm}
\usepackage{multicol}
\usepackage{multirow}
\usepackage{graphicx}
\usepackage{float}
\usepackage[english]{babel}
\usepackage{xcolor}
\newtheorem{exmp}{Example}
\usepackage[singlelinecheck=false]{caption}
\usepackage{subcaption}
\makeatletter
\renewcommand{\subsection}{%
	\@startsection{subsection}
	{2}
	{\z@}
	{-21dd plus-8pt minus-4pt}
	{10.5dd}
	{\normalsize\bfseries\boldmath}%
}
\makeatother
\usepackage{etoolbox}
\makeatletter
\patchcmd{\NAT@citex}
{\@citea\NAT@hyper@{%
		\NAT@nmfmt{\NAT@nm}%
		\hyper@natlinkbreak{\NAT@aysep\NAT@spacechar}{\@citeb\@extra@b@citeb}%
		\NAT@date}}
{\@citea\NAT@nmfmt{\NAT@nm}%
	\NAT@aysep\NAT@spacechar\NAT@hyper@{\NAT@date}}{}{}
\patchcmd{\NAT@citex}
{\@citea\NAT@hyper@{%
		\NAT@nmfmt{\NAT@nm}%
		\hyper@natlinkbreak{\NAT@spacechar\NAT@@open\if*#1*\else#1\NAT@spacechar\fi}%
		{\@citeb\@extra@b@citeb}%
		\NAT@date}}
{\@citea\NAT@nmfmt{\NAT@nm}%
	\NAT@spacechar\NAT@@open\if*#1*\else#1\NAT@spacechar\fi\NAT@hyper@{\NAT@date}}
{}{}
\begin{document}
	
	\title{The Shape of Phylogenies Under Phase--Type Distributed Times to Speciation and Extinction} 
	
	\titlerunning{PH Distributed Times to Speciation and Extinction}        
	
	\author{Albert Ch. Soewongsono$^{1}$\and Barbara R. Holland$^{1}$ \and Ma{\l}gorzata M. O'Reilly$^{1}$ 
	}
	
	
	\institute{Albert Ch. Soewongsono \at
		\email{albert.soewongsono@utas.edu.au}           
		\and
		Barbara R. Holland \at
		\email{barbara.holland@utas.edu.au}
		\and
		Ma{\l}gorzata M. O'Reilly \at
		\email{ma{\l}gorzata.oreilly@utas.edu.au}
		\and
		$^{1}$ School of Natural Sciences (Discipline of Mathematics), University of Tasmania, Hobart, 7005, Australia
	}
	\date{Received: date / Accepted: date}
	\maketitle		
	\begin{abstract}
		{Phylogenetic trees are widely used to understand the evolutionary history of organisms. Tree shapes provide information about macroevolutionary processes. However, macroevolutionary models are unreliable for inferring the true processes underlying empirical trees. Here, we propose a flexible and biologically plausible macroevolutionary model for phylogenetic trees where times to speciation or extinction events are drawn from a Coxian phase-type (PH) distribution. First, we show that different choices of parameters in our model lead to a range of tree balances as measured by Aldous' $\beta$ statistic. In particular, we demonstrate that it is possible to find parameters that correspond well to empirical tree balance. Next, we provide a natural extension of the $\beta$ statistic to sets of trees. This extension produces less biased estimates of $\beta$ compared to using the median $\beta$ values from individual trees. Furthermore, we derive a likelihood expression for the probability of observing any tree with branch lengths under a model with speciation but no extinction. 
			
			Finally, we illustrate the application of our model by performing both absolute and relative goodness-of-fit tests for two large empirical phylogenies (squamates and angiosperms) that compare models with Coxian PH distributed times to speciation with models that assume exponential or Weibull distributed waiting times. In our numerical analysis, we found that, in most cases, models assuming a Coxian PH distribution provided the best fit.}
	\end{abstract}
	
	\section{Introduction}
	\label{intro}
	Understanding how biodiversity is maintained and changed throughout time has been of long-standing interest in evolutionary biology \citep{quental2010diversity,morlon2014phylogenetic}. Fossil records are commonly used to make inferences about changes through time in speciation and extinction rates \citep{simpson1944tempo,stanley1998macroevolution,morlon2011reconciling}. However, most clades do not possess sufficiently complete fossil records to make such inferences \citep{ricklefs2007estimating,quental2010diversity}. In contrast, dated molecular trees are increasingly available, nevertheless, these ``reconstructed phylogenies" only give relationships between extant species \citep{nee1992tempo,nee1994extinction,stadler2013recovering}. These reconstructed phylogenies can also be used to study how diversification processes change throughout time \citep{nee1994extinction}, although some have argued that the use of reconstructed phylogenies needs to be accompanied with availability of fossil records \citep{quental2010diversity,morlon2014phylogenetic}. However, reconstructed phylogenies remain useful to study diversification and diversity dynamics when accompanied by biologically well-justified constraints \citep{louca2020extant}. Given the fact that we can use reconstructed phylogenies to understand about macroevolutionary processes, we then need mathematical models that are able to generate such trees.
	
	Several mathematical models have been proposed for studying macroevolutionary processes. These range from the constant-rate birth and death (crBD) model where speciation and extinction rates are assumed to be constant through time \citep{nee1994reconstructed}, to models where speciation and extinction rates change according to species age \citep{Hagen2015}, to models where an evolving trait can affect speciation and extinction rates \citep{maddison2007estimating,fitzjohn2012diversitree}. In addition, for models under the general birth-death process in which speciation and extinction rates can vary over time, a recent paper by \citet{louca2020extant} suggests that many of such models are congruent, meaning that these models generate the same expected lineage-through-time (LTT) plot. However, those models have different trends through time for their speciation and extinction rates. Given a choice of a model, various methods can be applied to use empirical data such as branch lengths from reconstructed trees to estimate the parameters of the model.
	These fitted parameters provide insight into speciation and extinction rates or structure of relationships between species through time \citep{harvey1991comparative,stadler2013recovering}. 
	
	In practice, given empirical data such as branch lengths from reconstructed phylogenies and a model, it is possible to derive an expression for the likelihood of observing these branch lengths and find the best-fitting parameters of the model using maximum-likelihood estimation (MLE) to make inference about the speciation and extinction rates \citep{morlon2011reconciling}. In order to see which model fits empirical data best, we can assess models via the likelihood ratio test (LRT) and the Akaike's Information Criterion (AIC) \citep{anderson2004model} or via the comparison of their simulated LTT plot, which counts the number of species that existed at each given time in the past, with an empirical LTT plot \citep{morlon2014phylogenetic}. Then, given a model with best choice of parameters, we can assess whether it fits well to the empirical data by comparing tree balance or tree topology and branch length distributions from empirical and simulated trees generated from the model. 
	
	The balance of a phylogenetic tree describes the branching pattern of the tree, ranging from imbalanced shape where sister clades tend to be very different in sizes to balanced shape where the clades are of similar sizes. Tree balance is important for understanding macroevolutionary dynamics on a tree \citep{Hagen2015} as it gives indication of heterogeneity of diversification rate across the tree without requiring information on branch lengths. To assess tree balance, summary statistics are used, and one that has been used to assess the goodness-of-fit of different macroevolutionary models to empirical data is Aldous' $\beta$ statistic. Many models in phylogenetics fail to resemble empirical datasets which often have $\beta$ value around -1 \cite{aldous1996probability}. For example, the simplest macroevolutionary model is the pure birth model, also known as the Yule-Harding (YH) model \citep{yule1925ii}, where each species is equally likely to speciate. It has been shown that trees under this model have the expected value $\beta=0$ In other words, the YH model predicts trees that are too balanced compared to empirical data \citep{aldous1996probability,aldous2001}. Likewise, models that include diversity-dependent \citep{etienne2012diversity} and time-dependent speciation and extinction have been shown to produce the same expected tree balance as the YH model \citep{lambert2013birth}. These models fall under a general class of species-speciation-exchangeable models as described in \cite{stadler2013recovering}. This suggests that this class of models is not adequate to explain the macroevolutionary dynamics that has produced empirical trees, thus  better models are required. 
	
	Next, if we consider branch lengths, then one can use the $\gamma$ statistic \citep{pybus2000} to compare model-generated phylogenies with empirical data. This summary statistic is used for determining whether there has been increase or decrease in diversification rate (speciation rate minus extinction rate) through time. It has been shown that empirical $\gamma$ value tend to be below 0, which indicates a slowdown in diversification rate \citep{phillimore2008density,rabosky2008density,morlon2010inferring}.
	
\begin{figure}[!htbp]
		\centering
		\includegraphics[scale=0.28]{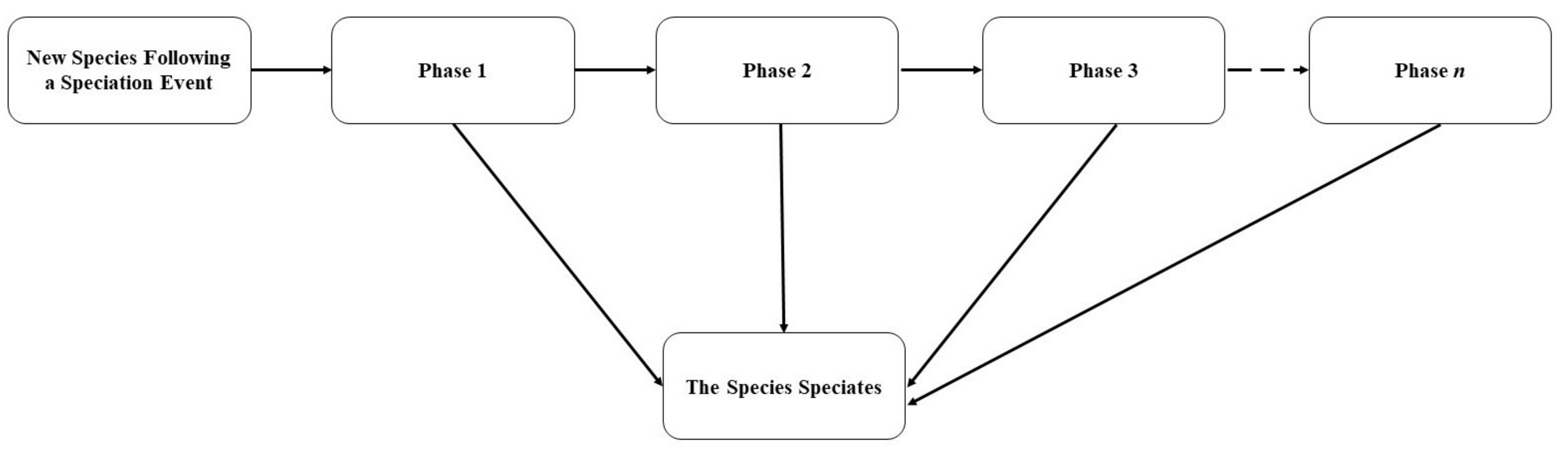}
		\nobreak
		\caption{A new species passes through different phases during its `lifetime' until the next speciation event. Each phase corresponds to a non-absorbing state in a CTMC and speciation corresponds to the single absorbing state. At the start, the species directly goes to phase 1 where it can either undergo speciation or move to the next phase with certain rates. The process can continue up to a finite number of $n$ phases, each corresponding to a different rate of speciation}
		\label{speciesphasetype}
	\end{figure}

	In this paper, we construct a stochastic model for generating species phylogenies in which we apply Coxian PH distributions \citep{neuts1981,marshall2004using} for times to speciation and times to extinction. PH distributions describe the time to absorption in a continuous-time Markov chain (CTMC) with a single absorbing state and a finite number of non-absorbing states. Biologically this could be thought of as a species passing through different phases where it may be more or less likely to speciate depending on a current underlying phase (Fig.~\ref{speciesphasetype}). Similarly, times to extinction can also be modelled using phase-type distributions.
	
	Next, Assuming that speciation is symmetric \citep{stadler2013recovering}, then whenever speciation occurs, two new independent PH distributions are initiated for the descendant species.
	
	Below, we show that different parameter choices for age-dependent speciation rates produce phylogenetic trees that can range from highly balanced to highly unbalanced. In particular, we find parameters that give similar tree balance statistics to empirical trees. Similar to the results in \cite{Hagen2015}, we observe that different parameter choices for age-dependent extinction rates produce a range of diversification dynamics, as measured by the $\gamma$ statistic. 
	
	Furthermore, we develop a new approach to compute the $\beta$ statistic by considering a set of trees instead of computing $\beta$ from a single tree and then taking an average. In our simulations, we found that this approach leads to more accurate results in computing the $\beta$ statistic, particularly for trees with fewer extant species.
	
	In order to illustrate the application of our methodology, we consider a special case of our model in which only speciation (and not extinction) occurs. We derive a likelihood expression to compute the probability of observing any tree with branch lengths, under such model. Next, using the squamate reconstructed phylogeny \citep{pyron2013phylogeny} and the angiosperm reconstructed phylogeny \citep{zanne2014three}, we perform model selection across different clades on both trees to compare our model to models that assume an exponential or Weibull distribution for the speciation process. In our numerical analysis, in which we used branch lengths from various clades in empirical phylogenies as our data, we found that using a Coxian PH distribution for times to speciation provided a better fit to the empirical data compared to either the exponential or Weibull distribution.
	
	The rest of our paper is structured as follows. In the methods section below, we first summarise the key properties of the PH distribution and construct two examples of Coxian PH distributions to be applied in our analysis. Next, we present our method for calculating the $\beta$ statistic for a set of trees, derive a likelihood expression based on our model for fitting empirical tree data, and describe our simulation studies. In the results section that follows, we discuss the simulations and the application of our model to empirical trees. In summary, we found that Coxian PH distributions are a useful tool for studying macroevolutionary dynamics.
	
	\section{Methods}
	\label{sec:1}
	\subsection{PH Distribution and Relevant Properties}
	\label{sec:2}
	In this section we introduce the PH distribution and some of its key properties.
	\begin{definition}\textbf{(Continuous PH distributions)} Let $\{X(t) : t \geq 0\}$ be a continuous time Markov chain defined on state space $S = \hat{S} \bigcup \{n+1\}$, where $\hat{S}=\{1,2,\cdots,n\}$ is the set of non-absorbing states and $n+1$ is an absorbing state, initial distribution vector $\bm{\alpha}=[\alpha_{i}]_{i \in \hat{S}}$, and generator matrix 
		\begin{eqnarray}
			\mathbf{Q^{*}} 
			= [Q^{*}_{i,j}]_{i,j \in S}
			= \begin{bmatrix}
				\mathbf{Q} & \mathbf{q}\\
				\mathbf{0} &0 
			\end{bmatrix},
		\end{eqnarray}
		where $\mathbf{Q}=[Q_{i,j}]_{i,j\in \hat{S}}$ is a square matrix with dimension $n$ that records the transition rates between non-absorbing states $i,j\in \hat{S}$, $\mathbf{q}=[Q_{i,n+1}]_{i\in \hat{S}}$ is a column vector that records the transition rates from non-absorbing states $i \in \hat{S}$ to the absorbing state $n+1$, and $\mathbf{0}$ is the row vector with corresponding dimension. By the definition of generator matrix $\mathbf{Q}$, we have $Q_{i,i} < 0,\text{for all } i$, $Q_{i,j} \geq 0$ for $i \neq j$, and $\mathbf{Q}\mathbf{1}+\mathbf{q}=\mathbf{0}$, where $\mathbf{q}$ is the exit rate vector.

		Let $Z=\text{inf}\left\{t \geq 0 : X(t)=n+1\right\}$ be the random variable recording the time until absorption, then $Z$ is said to be continuous PH distributed with parameters $\bm{\alpha}$ and $\mathbf{Q}$, which we denote $Z \sim PH\left(\bm{\alpha},\mathbf{Q}\right)$.
	\end{definition}
	
	\begin{theorem}{\textit{(The cumulative distribution and density functions of continuous PH distribution)}} Suppose $Z \sim PH_{c}\left(\bm{\alpha},\mathbf{Q}\right)$, then the cumulative distribution and the probability density function of $Z$ are given respectively by,
		\begin{eqnarray}
			&F_{Z}(z) = 1-\bm{\alpha}e^{\mathbf{Q}z}\mathbf{1},\\
			&f_{Z}(z) = 
			\bm{\alpha}e^{\mathbf{Q}z}\mathbf{t},
		\end{eqnarray}
		and its mean and variance are given by,
		\begin{eqnarray}
			&E(Z)=	-\bm{\alpha}\mathbf{Q}^{-1}\mathbf{1},\\
			&Var(Z)= 2\bm{\alpha}\mathbf{Q}^{-2}\mathbf{1}-\left(\bm{\alpha}\mathbf{Q}^{-1}\mathbf{1}\right)^{2}. 
		\end{eqnarray}
		Proof of this theorem is originally given in \cite{neuts1975probability}, and a clear exposition is given in \cite{verbelen2013}.\qed
	\end{theorem}
	
	\begin{definition}\textbf{(Coxian PH distribution)}
		\label{generalCox} If $\bm{\alpha}$ and $\mathbf{Q}$ are defined as,
		\begin{eqnarray}
			&\bm{\alpha} &= [1,0,\cdots,0],\\
			&\mathbf{Q}&=\begin{bmatrix}
				-\lambda_{1} & p_{1}\lambda_{1}& 0 & \dots & 0 & 0\\
				0 & -\lambda_{2} & p_{2}\lambda_{2} & \ddots & 0 & 0 \\
				\vdots & \ddots & \ddots & \ddots & \ddots & \vdots\\
				0 & 0 & \ddots & -\lambda_{n-2} & p_{n-2}\lambda_{n-2} & 0\\
				0 & 0 & \dots & 0 & -\lambda_{n-1} & p_{n-1}\lambda_{n-1} \\
				0 & 0 & \dots & 0 & 0 & -\lambda_{n}
			\end{bmatrix},
		\end{eqnarray}
		where $0 < p_{i} \leq 1$ and $\lambda_{1},\cdots,\lambda_{n}>0$ for all $i = 1,2,\dots n-1$, then we say that the random variable $T \sim PH\left(\bm{\alpha},\mathbf{Q}\right)$ follows Coxian PH distribution.
	\end{definition}
	
	\cite{cumani1982canonical} showed that any acyclic PH (APH) distribution (including Coxian PH distributions), that is, a distribution with an upper triangular generator matrix \citep{asmussen1996}, can be restructured to a canonical form such as shown above, and thus only requires $2n$ parameters as opposed to $n^{2}+n$ parameters for a general PH distribution. This reduction in the number of parameters makes it computationally simpler to fit parameters \citep{thummler2006novel}. Further, \cite{cumani1982canonical} and \cite{dehon1982geometric} showed that for any APH distribution, there exists an equivalent representation as a Coxian PH distribution with $\lambda_{1}\leq \lambda_{2} \leq \cdots \leq \lambda_{n}$.\\
	
	In the results section later on, we use branch lengths from data to find parameters from a PH distribution that provide best fit to the data. Thus, it is necessary to fix the number of non-absorbing states. \cite{thummler2006novel} stated that it is not appropriate to fit general PH distributions if the number of non-absorbing states is above four, due to high computational cost and dependence on the initial values. Therefore, in this paper, we consider PH distributions with four non-absorbing states and an absorbing state. We note that in our analysis we found that having more than four non-absorbing states, did not significantly change the shape of the distribution fitted to the simulated data, or the quality of the numerical results. 
	
	\subsection{Coxian-Based Macro-Evolutionary Model}
	\label{sec:3}
	Now, we develop a stochastic model for generating species phylogenies, in which we assume that the time spent by each newly formed lineage before the next speciation or extinction event is drawn from a Coxian PH distribution. Our model is a special case of the well-studied Bellman-Harris model which allows any distribution of waiting times to extinction or speciation \citep{bellman1948theory}.
		This model is discussed in \cite{hagen2018treesimgm} and they provide an \textit{R} package \citep{hagen2018treesimgm} that allows users to simulate trees under a general Bellman Harris model. However, while it is possible to simulate trees under this very general class of models it is not possible to fit parameters of a general Bellman-Harris distribution to empirical data. 
		A novelty of our approach is that we are able derive a likelihood expression for the probability of observing a reconstructed phylogeny under our model in the case with no extinction, and that we can therefore fit parameters. We show that the model fits empirical data better than other models, including the Weibull-based age-dependent speciation model of \cite{Hagen2015}, which is also a special case of the Bellman-Harris model.
	
	In our model, we focus on speciation event that follows symmetric process where neither child lineage inherits its parent's age. Thus, each branch length on a given tree can be thought of as an independent random variable drawn from the imposed Coxian PH distribution.
	
	Also, we construct two examples of Coxian PH distributions that can be applied to either the speciation or extinction process. One is such that the rate of absorption increases for each non-absorbing state, and the other is such that the rate of absorption decreases for each non-absorbing state. In biological terms, this choice of rates corresponds to an increasing or decaying rate of speciation or extinction. 
	
	In the results section, we show that different parameterizations of these two examples provide a sufficient range of mean and second moment values to approximately match the mean and second moment values from other known distributions or from data (see the appendix). In each example, we assume that the time $T$ to speciation or extinction follows a Coxian PH distribution with parameters $\bm{\alpha}=\left[1,0,0,0\right]$ and $\mathbf{Q}$, and write $T \sim PH(\bm{\alpha},\mathbf{Q})$, which corresponds to an absorbing Markov chain with four non-absorbing states and an absorbing state.
	
	\begin{exmp}\textbf{(Coxian PH Distributed Model for Decreasing Rate)}\label{model}
		\begin{align}
			&\mathbf{Q}=\begin{bmatrix}
				-z & (1-y)z & 0 & 0 & \\
				0 & -(1+x) & \left(1-y^{2}\right)(1+x) & 0  \\
				0 & 0 & -\left(1+x^{2}\right) & \left(1-y^{3}\right)\left(1+x^{2}\right) \\
				0& 0 & 0 & -x^{3}
			\end{bmatrix}
			\text{, }\bm{q}=\begin{bmatrix}
				yz \\
				y^{2}(1+x) \\
				y^{3}\left(1+x^{2}\right) \\
				x^{3}
			\end{bmatrix},
		\end{align}
		where $0<x\leq 1$, $0<y<1$, $z \geq 2$ and $\bm{q}$ is the exit rate vector. 
	\end{exmp}
	
	\begin{exmp}\textbf{(Coxian PH Distributed Model for Increasing Rate)}\label{model2}
		\begin{align}
			&\mathbf{Q}=\begin{bmatrix}
				-\left(1+x^3\right) & \left(1-y^4\right)\left(1+x^3\right) & 0 & 0 & \\
				0 & -\left(1+x^2\right) & \left(1-y^{3}\right)\left(1+x^2\right) & 0  \\
				0 & 0 & -(1+x) & \left(1-y^{2}\right)(1+x) \\
				0& 0 & 0 & -z
			\end{bmatrix}
			\text{, }\bm{q}=\begin{bmatrix}
				y^{4}\left(1+x^{3}\right) \\
				y^{3}\left(1+x^{2}\right) \\
				y^{2}\left(1+x\right) \\
				z
			\end{bmatrix},
		\end{align}
		where $0<x\leq 1$, $0<y<1$, $z \geq 2$ and $\bm{q}$ is the exit rate vector.
	\end{exmp}
	
	By standard theory of the PH distribution, the first and second moments of the Coxian PH distribution in Example~\ref{model} and Example~\ref{model2} are given by,
	\begin{eqnarray}
		E_{PH}(X) &=&\frac{1}{z}+(1-y)\left(\frac{1}{1+x}+\left(1-y^2\right)\left(\frac{1}{1+x^2}+\frac{1-y^3}{x^3}\right)\right), \nonumber \\
		E_{PH}\left(X^{2}\right)&=& \frac{2}{z^{2}}+\frac{2(1-y)}{1+x}\left(\frac{1}{z}+\frac{1}{1+x}\right)+\frac{2(1-y)\left(1-y^2\right)}{1+x^2}\left(\frac{1}{z}+\frac{1}{1+x}+\frac{1}{1+x^2}\right)\nonumber \\
		&&
		+\frac{2(1-y)\left(1-y^2\right)\left(1-y^3\right)}{x^3}\left(\frac{1}{z}+\frac{1}{1+x}+\frac{1}{1+x^2}+\frac{1}{x^3}\right),
		\label{moments1}
	\end{eqnarray}
	and
	\begin{eqnarray}
		E_{PH}(X) &=&\frac{1}{1+x^{3}}+\left(1-y^{4}\right)\left(\frac{1}{1+x^{2}}+\left(1-y^3\right)\left(\frac{1}{1+x}+\frac{1-y^2}{z}\right)\right),\nonumber \\
		E_{PH}\left(X^{2}\right)&=& \frac{2}{\left(1+x^3\right)^2}+\frac{2\left(1-y^4\right)}{1+x^2}\left(\frac{1}{1+x^3}+\frac{1}{1+x^2}\right)+\frac{2\left(1-y^4\right)\left(1-y^3\right)}{1+x}\nonumber \\
		&&\left(\frac{1}{1+x^3}+\frac{1}{1+x^2}+\frac{1}{1+x}\right) +\frac{2\left(1-y^4\right)\left(1-y^3\right)\left(1-y^2\right)}{z}\nonumber\\
		&&\left(\frac{1}{1+x^3}+\frac{1}{1+x^2}+\frac{1}{1+x}+\frac{1}{z}\right),
		\label{moments2}
	\end{eqnarray}
	respectively. The derivations of Eq.~\ref{moments1} and Eq.~\ref{moments2} are shown in the Appendix.
	
	\subsection{Novel Approach for Computing $\beta$ Values}
	\label{sec:4}
	Here, we propose the following new approach for estimating the $\beta$ value from a set of trees $\{T_1,\ldots, T_M\}$, which can be either empirical trees or simulated trees under some model of interest, which consists of the following two steps.

	In the first step, for each individual tree $T_m$, $m=1,\ldots ,M$, we compute the probability $q_{n}(i,\beta)$ of observing $i$ left tips from each split on a tree with $n$ extant tips, using Eq.~{4} in \cite{aldous1996probability},
	\begin{equation}
		q_{n}(i,\beta) = \frac{1}{a_{n}(\beta)}\frac{\tau(\beta+i+1)\tau(\beta+n-i+1)}{\tau(i+1)\tau(n-i+1)},1 \leq i \leq n-1,
		\label{qn}
	\end{equation}
	stated for the Aldous' $\beta$-splitting model in~\cite{aldous1996probability}, where $a_{n}(\beta)$ is the normalizing constant. In the case where the tree size is too large, the above expression is not numerically tractable. So, we use the following approximation instead, which is also used in the apTreeShape package \citep{bortolussi2006aptreeshape}, given by,
	\begin{equation}
		q_{n}(i,\beta) = \frac{1}{\hat{a}_{n}(\beta)}\left(\frac{i}{n}\right)^{\beta}\left(1-\frac{i}{n}\right)^{\beta},
		\label{qaltn} 
	\end{equation}
	where $\hat{a}_{n}(\beta)$ is the normalizing constant. Proof of Eq.~\ref{qaltn} is summarised in the Appendix.
	
	In the second step, we find $\beta$ which maximises the product of all probabilities in Eq.~\ref{qn} (or in~\ref{qaltn} if more suitable), over {\em all splits of all trees} in the tree set $\{T_1,\ldots, T_M\}$.
	
	This approach of computing a $\beta$ value for a set of trees is useful in the context of simulated tree data -- we will show that it gives less biased results than averaging $\beta$ values over sets of trees. Beyond simulation studies, there may be other contexts where it is useful to estimate $\beta$ for a set of trees. For example, when studying bio--geographic patterns researchers may  have multiple species trees for the same set of geographic regions. It would also be possible to compute a single $\beta$ value for a set of gene trees.   
	
	Below, we illustrate the above procedure by generating a set of trees $\{T_1,\ldots, T_M\}$ under a pure-birth process where times to speciation were drawn from a Coxian PH distribution in Definition~\ref{generalCox} with parameters
	\begin{equation}
		\bm{\alpha} = [1,0,0,0], \quad
		\mathbf{Q}=\begin{bmatrix}
			-\lambda_{1} & p_{1}\lambda_{1} & 0 & 0 & \\
			0 & -\lambda_{2} & p_{2}\lambda_{2} & 0 & \\
			0 & 0 & -\lambda_{3} & p_{3}\lambda_{3} \\
			0 & 0 & 0 & -\lambda_{4}
		\end{bmatrix},
	\end{equation}
	where $\lambda_{1}=2,\lambda_{2}=1.1,\lambda_{3}=1.01,\lambda_{4}=0.001,p_{1}\lambda_{1}=p_{2}\lambda_{2}=p_{3}\lambda_{3}=1$, which gives
	\begin{equation}
		\mathbf{Q}=\begin{bmatrix}
			-2 & 1 & 0 & 0 & \\
			0 & -1.1 & 1 & 0 & \\
			0 & 0 & -1.01 & 1 \\
			0 & 0 & 0 & -0.001
		\end{bmatrix}, \quad  
		\bm{q} =
		\begin{bmatrix}
			1 \\
			0.1 \\
			0.01\\
			0.001\\
		\end{bmatrix}.
		\label{simplecox}
	\end{equation}
	We note that the structure of the exit rate vector $\bm{q}$ implies that the probability of getting absorbed from later states is less likely than from earlier states.
	
	In our illustration, we limit the range of $\beta$ to the interval between $-2$ and $10$, and only consider splits when the number of tips is greater than or equal to $4$. We do so due to the fact that for $3$ or fewer species, we can only have one combination of the number of tips on the left and on the right, which is not informative.
	
	\subsection{Likelihood-Based Inference using Empirical Branch Length Data}
	\label{sec:5}
	In this section, we propose a method for finding parameters of a PH distribution using branch length data from a phylogenetic tree. We assume that the time until a speciation event on a branch follows a PH distribution and that there is no extinction. We write the likelihood expression using parameters from the PH distribution to calculate the probability of observing a tree with a given number of extant species. 
	
\begin{figure}[!htbp]
		\centering
		\includegraphics[scale=0.5]{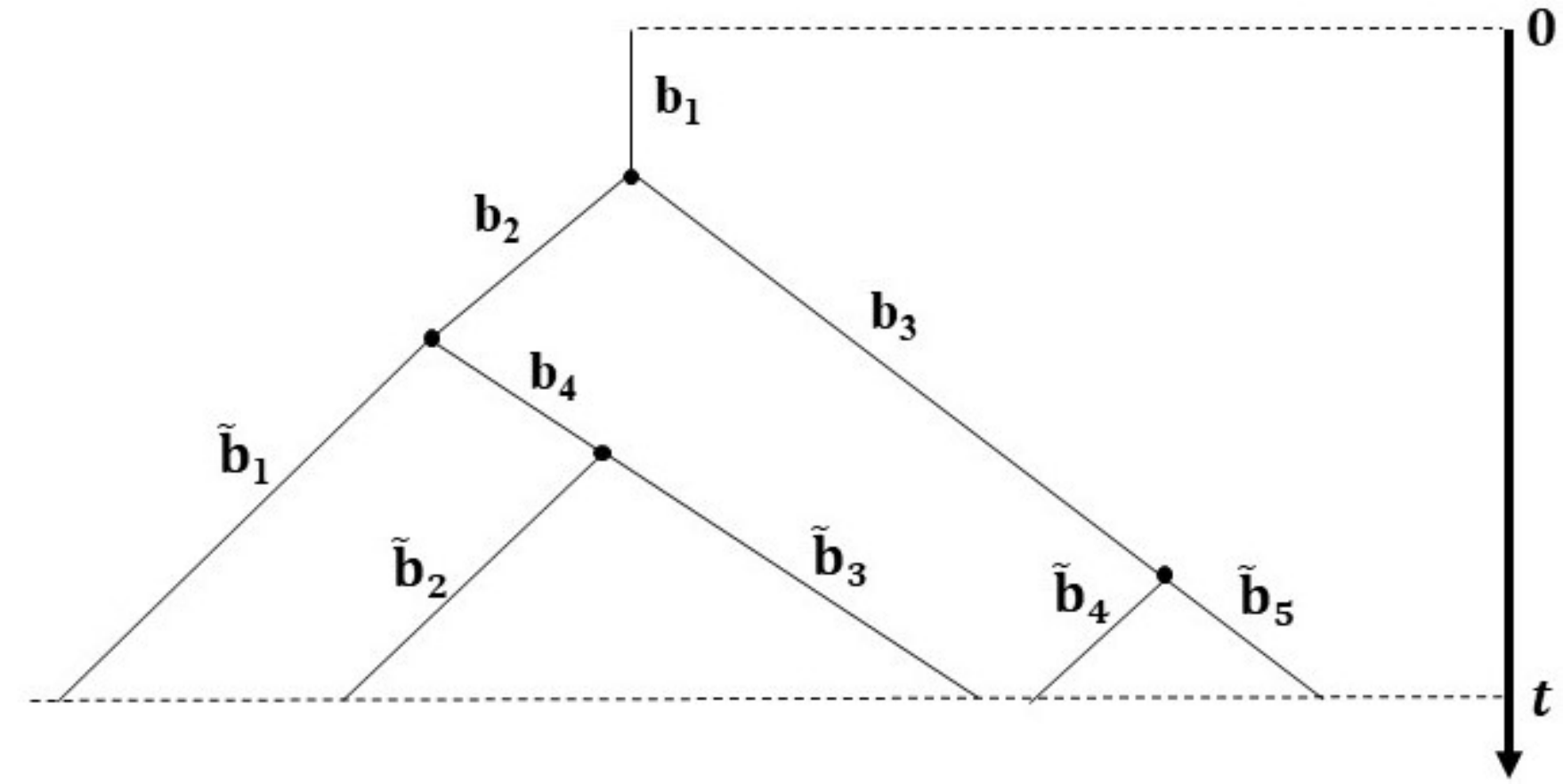}
		\nobreak
		\caption{Phylogenetic tree with five extant tips evolving under a symmetric speciation mode. Branch lengths are independent and drawn from the same PH distribution $PH\left(\bm{\alpha},\mathbf{Q}\right)$}
		\label{probtree}
	\end{figure}

	Assuming that a tree evolves under a symmetric speciation mode, and that times to speciation events are drawn from a PH distribution with some parameters, we can treat each branch length on the tree as independently drawn from the same PH distribution. We illustrate this in Fig.~\ref{probtree}, in which the lengths of internal branches and pendant branches are denoted by $\{b_{1},b_{2},b_{3},b_{4}\}$ and  $\{\tilde{b}_{1},\tilde{b}_{2},\tilde{b}_{3},\tilde{b}_{4},\tilde{b}_{5}\}$, respectively. 
	
	In general, we denote the lengths of internal and pendant branches by $b_i$, for $i=1,\ldots,k$, and $\tilde b_j$, for $j=1,\ldots,\ell$, where the total number of internal branches and pendant branches is denoted by $k$ and $\ell$, respectively. Here, because we consider the root branch, we note that $k=\ell-1$. Both internal and pendant branches follow a PH distribution with parameter $\bm{\alpha}$ and rate matrix $\mathbf{Q}$, that is, $b_i,\tilde b_j \sim PH\left(\bm{\alpha},\mathbf{Q}\right)$. It follows from the properties of the PH distribution~\citep{neuts1981},  that the probability of observing an internal branch of length $b_i$ is the probability density of the distribution along the branch given by $\bm{\alpha}e^{\mathbf{Q}b_{i}}\bm{q}$ and the probability of observing a pendant branch of length $\tilde b_j$ is the probability that the branch has survived until time t (i.e. the cumulative probability of the distribution) given by $\bm{\alpha}e^{\mathbf{Q}\tilde b_j}\bm{1}$, where $\bm{1}$ is a column vector of ones. Therefore, by independence of the branch lengths, the probability of observing tree $T$ can be written as,
	\begin{align}
		\text{Pr}(T) = \prod_{i=1}^{k}\left(\bm{\alpha}e^{\mathbf{Q}b_{i}}\bm{q}\right) \times \prod_{j=1}^{\ell}\left(\bm{\alpha}e^{\mathbf{Q}\tilde b_{j}}\bm{1}\right),
		\label{prob_treenoext}
	\end{align}
	with $\bm{\alpha} = [1,0,\cdots,0]$, since we apply Coxian PH distribution.
	
	Given the branch lengths of a {\em single tree} $T$, we apply the standard procedure of maximum likelihood estimation which maximises the probability in Eq.~\ref{prob_treenoext}, to find the best-fitting parameter $\mathbf{Q}$ of the Coxian PH distribution. Alternatively, given the branch lengths of a {\em tree set} $\{T_1,\ldots, T_M\}$, we apply maximum likelihood estimation  to maximise the product
	\begin{align}
		\text{Pr}(\{T_1,\ldots, T_M\}) = 
		\text{Pr}(T_1)\cdots \text{Pr}(T_M),
		\label{prob_treeset}
	\end{align}
	where we assume tree are independent and apply Eq.~\ref{prob_treenoext} to compute the probabilities of observing the individual trees $T_1,\ldots, T_M$.
	
	\subsection{Simulations}
	\label{sec:6}
	To compare $\beta$ values estimated from an individual tree or a tree set, we performed the following analysis.
	\begin{itemize}
		\item First, we simulated sets of $1000$ trees using \textit{TreeSimGM} package, where each tree from the same set had the same number of extant tips $n\in \{10,20,30,\cdots,200\}$ and their times to speciation event were drawn from PH distribution with rate matrix defined in Eq.~\ref{simplecox}. In order to validate our results, we also simulated sets of trees evolving under the YH model, since it is known that trees evolving under this model correspond to the value $\beta=0$ \citep{aldous2001}.

		\item Next, using maximum likelihood estimation, we computed the best-fitting $\beta$ statistic between $-2$ and $10$ to maximise the product of $q_{n}(i,\beta)$ probabilities in Eq.~\ref{qn}. Our custom R script, based on \textit{maxlik.betasplit} function from the apTreeShape package \citep{bortolussi2006aptreeshape} to estimate $\beta$ from sets of trees, is available as a Supplementary Material on Dryad.

		\item Finally, we computed the $95\%$ confidence intervals for the estimated $\beta$ values, denoted $\hat{\beta}$, from each tree set. In order to get the lower and upper bound for the confidence intervals, we performed a numerical search over $500$ equidistant points between $\hat{\beta}-5\times SE\left(\hat{\beta}\right)$ and  $\hat{\beta}$ to find the point that corresponds to the lower bound and $500$ equidistant points between $\hat{\beta}$ and $\hat{\beta}+5\times SE\left(\hat{\beta}\right)$ to find the point that corresponds to the upper bound. The lower and upper bounds were chosen such that their likelihood is equal to the likelihood of the MLE minus a half of the chi-square value with 1 degree of freedom; this give a 95\% confidence interval \citep{pawitan2001all}.
		The standard error for $\hat{\beta}$, $SE\left(\hat{\beta}\right)$, was evaluated using
		\begin{equation}
			SE\left(\hat{\beta}\right) = \frac{1}{\sqrt{I\left(\hat{\beta}\right)}},
		\end{equation}
		where $I\left(\hat{\beta}\right)$ is the Fisher information of $\hat{\beta}$.
	\end{itemize}
	
	Following this, we investigated whether speciation or extinction processes control tree balance, as measured by the $\beta$ statistic, or the change in diversification dynamics, as measured by the $\gamma$ statistic. We simulated trees under our model in Example~\ref{model} to study the effect of different choices of parameter values for speciation and extinction on tree balance and diversification dynamics, as follows.
	\begin{itemize}
		\item First, in Eq.~\ref{moments1}, we set $z=10$ and mean waiting time to both speciation and extinction $E_{PH}(X) = 2$.  The choice of $E_{PH}(X)$ scales the branch lengths of generated phylogenies, but results will be invariant to this choice of the mean since we only consider tree balance and relative branch lengths. Next, we searched for parameters $x$ and $y$ that correspond to the following coefficients of variation $CV = \frac{\sigma}{\mu} \in\{30.08,13.49,5.53,1.49\}$.
		
		\item Further, we simulated $300$ trees with $100$ extant tips under symmetric and asymmetric speciation modes, in which times to speciation followed a PH distribution with parameters $x$, $y$ and $z$ discussed above, while times to extinction followed an exponential distribution with some rate $\lambda$. Then we repeated this procedure, by assuming exponential distribution for the times to speciation, and PH distribution for the times to extinction. In our simulation, we applied the \textit{TreeSimGM}  package in \textit{R}~\citep{hagen2018treesimgm} to simulate reconstructed phylogenetic trees.

		\item We then studied the effect of the parameters in models for speciation and extinction, on the tree balance as measured by $\beta$ statistic. We computed the $\beta$ statistic in two different ways, by computing $\beta$ for each individual tree, using the \textit{apTreeshape} package \citep{bortolussi2006aptreeshape}, and from a set of trees based on our new approach as discussed earlier, using our code provided in the Supplementary Material available on Dryad. 
		
		\item Finally, we studied the effect of the parameters in models for speciation and extinction, on the branch lengths as measured by $\gamma$ statistic, using the \textit{APE} package \citep{paradis2004ape}.
	\end{itemize}

	We propose that in order to fit parameters of the PH distribution using branch length data, we apply the probability of observing the given {\em set of trees} given Eq.~\ref{prob_treenoext}-\ref{prob_treeset}, and the maximum likelihood estimation.
	
	To illustrate this method, we simulated trees for a range of distributions modelling the times to speciation (and without extinction events), and then fitted the parameters of the PH distribution of Example~\ref{model} to the obtained branch lengths data. In total, we generated $4,900$ branches in trees with $50$ tips each, using \textit{TreeSimGM} package. We then gathered internal branch lengths and pendant branch lengths from the generated trees as two separate sets. We computed the probability of observing the given set of trees using Eq.~\ref{prob_treenoext}-\ref{prob_treeset}. Then, we found parameters $x$, $y$, and $z$ via maximum likelihood estimation using the built-in \textit{R} function, \textit{optim}, under ``\textit{L-BFGS-B}" method \citep{byrd1995limited} with multiple starting points of $x,y,z$, followed by local optimisation using ``Nelder-Mead" method \citep{nelder1965simplex}. Also, we plotted the density of the fitted distribution and the known assumed distribution used to simulate the data. Finally, using the fitted parameters $x$, $y$ and $z$, we we generated trees with the same number of tips as in the simulated data, and compared their distribution of branch lengths with that of the simulated trees.

	To complete the analysis, we also applied the above method for parameter fitting under a model with speciation as well as extinction, assuming that times to speciation follow PH distribution while times to extinction follow an exponential distribution with some constant rate $\lambda$. We used $\lambda=0.1$ and $\lambda=0.4$ to study increasing bias from ignoring the extinction process in Eq.~\ref{prob_treenoext}.
	
	\subsection{Fitting to Empirical Data}
	\label{sec:7}
	Below, we fit the parameters of our macroevolutionary model without extinction to empirical phylogenies of squamates \citep{pyron2013phylogeny} and angiosperms \citep{zanne2014three}. We apply the following three variants of the model: a model with general PH distribution in Definition~\ref{generalCox}, and two models with Coxian PH distribution in Example~\ref{model} and Example~\ref{model2}, respectively. We also fit the parameters of two other models without extinction, in which the times to speciation event follow exponential distribution and Weibull distribution, respectively, to the same data. 
	
	We then select the model that is the best fit to the data, out of the above five alternatives. This task is performed using branch length data across different clades on the squamate and angiosperm phylogenies. The best model is chosen based on their Akaike's Information Criterion (AIC) values \citep{akaike1998information}.

	For each model, we simulated one tree corresponding to each clade, with the same number of extant tips in each clade. We then compared branch length distributions in the simulated trees to those of the empirical trees, and plotted the hazard function of speciation for each clade. In order to view these phylogenies and to extract the clades, we used Dendroscope 3 software \citep{huson2012dendroscope}.
	
	\section{Results}
	\label{sec:8}
	
	\subsection{Our Macroevolutionary Model Leads to a Wide Range of Tree Shapes}
	\label{sec:9}
	
\begin{figure}[!htbp]
		\centering
		\captionsetup{width=\linewidth}
		\begin{subfigure}[!htbp]{0.43\textwidth}
			\centering
			\includegraphics[scale = 0.3]{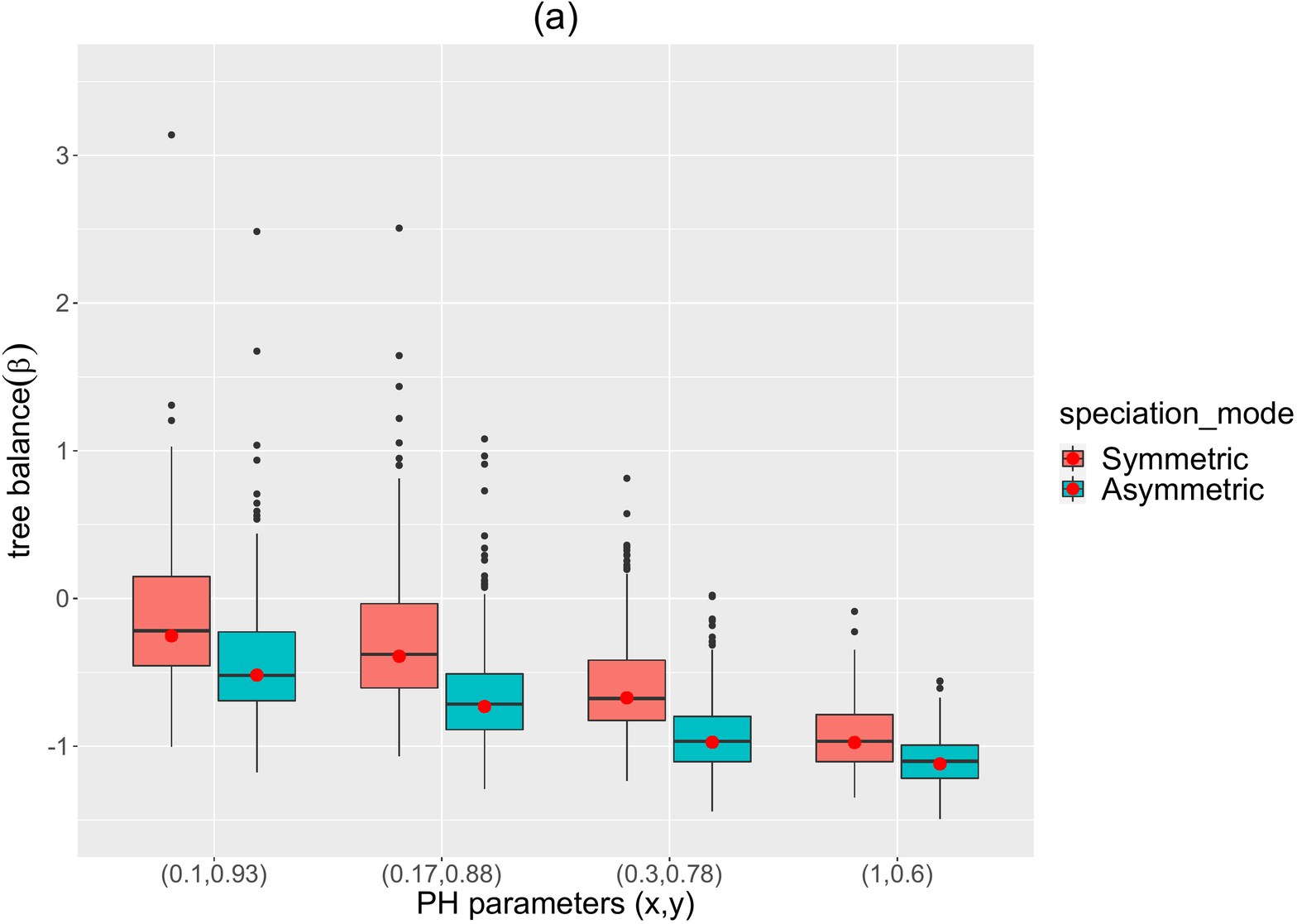}
		\end{subfigure}%
		~ 
		\begin{subfigure}[!htbp]{0.5\textwidth}
			\centering
			\includegraphics[scale = 0.3]{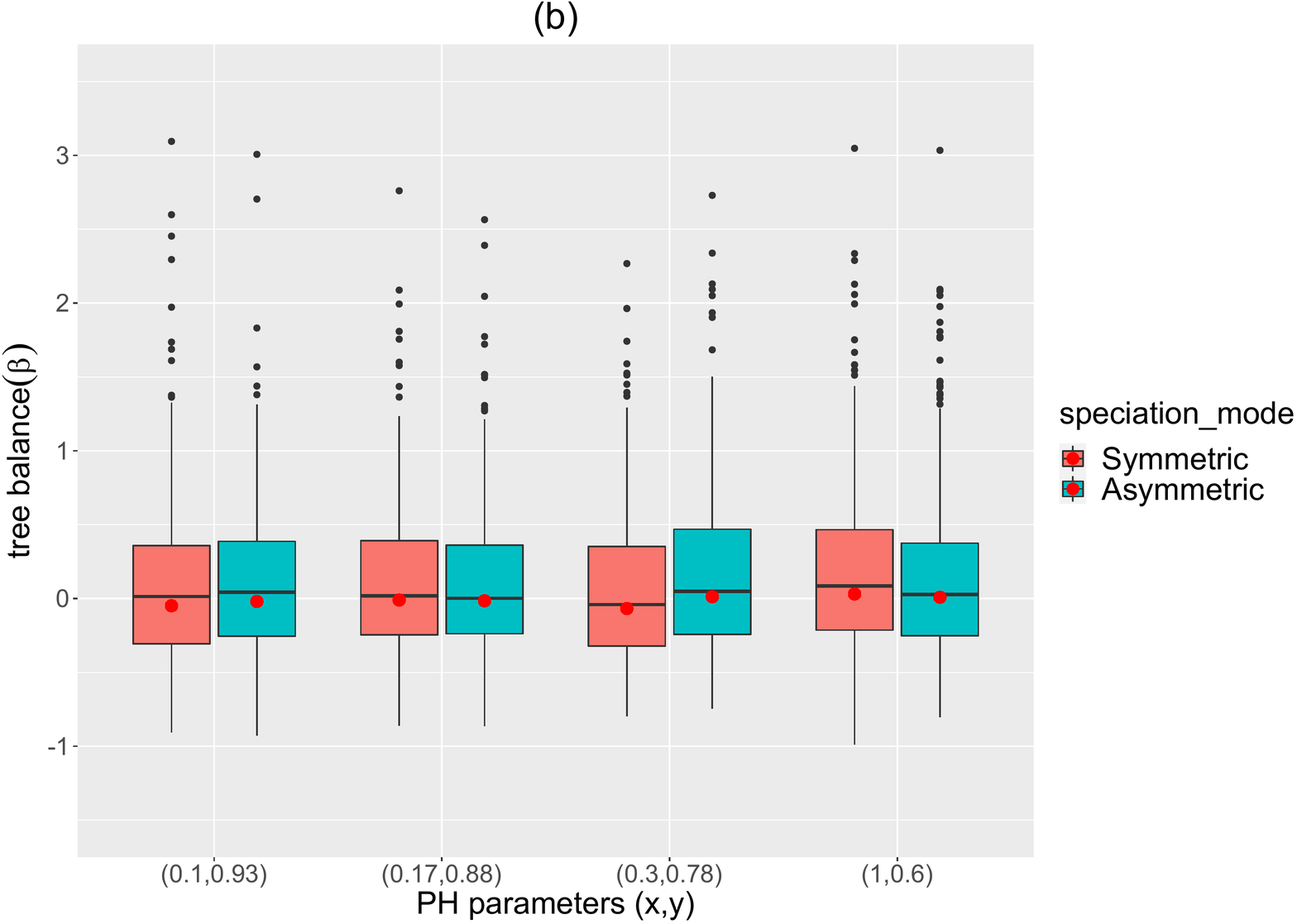}
		\end{subfigure}
		~ 
		\begin{subfigure}[!htbp]{0.43\textwidth}
			\centering
			\includegraphics[scale = 0.3]{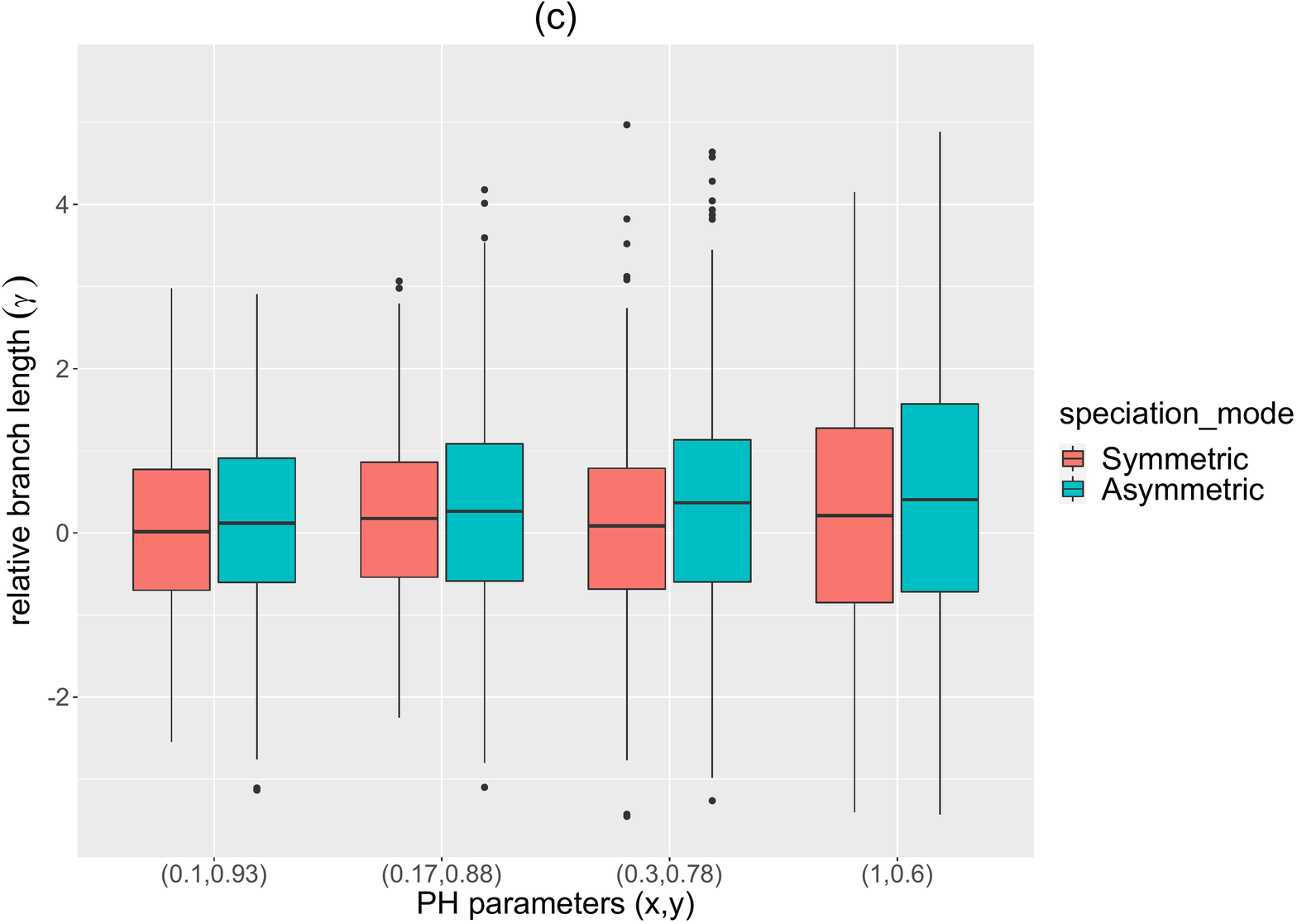}
		\end{subfigure}%
		~
		\begin{subfigure}[!htbp]{0.5\textwidth}
			\centering
			\includegraphics[scale = 0.3]{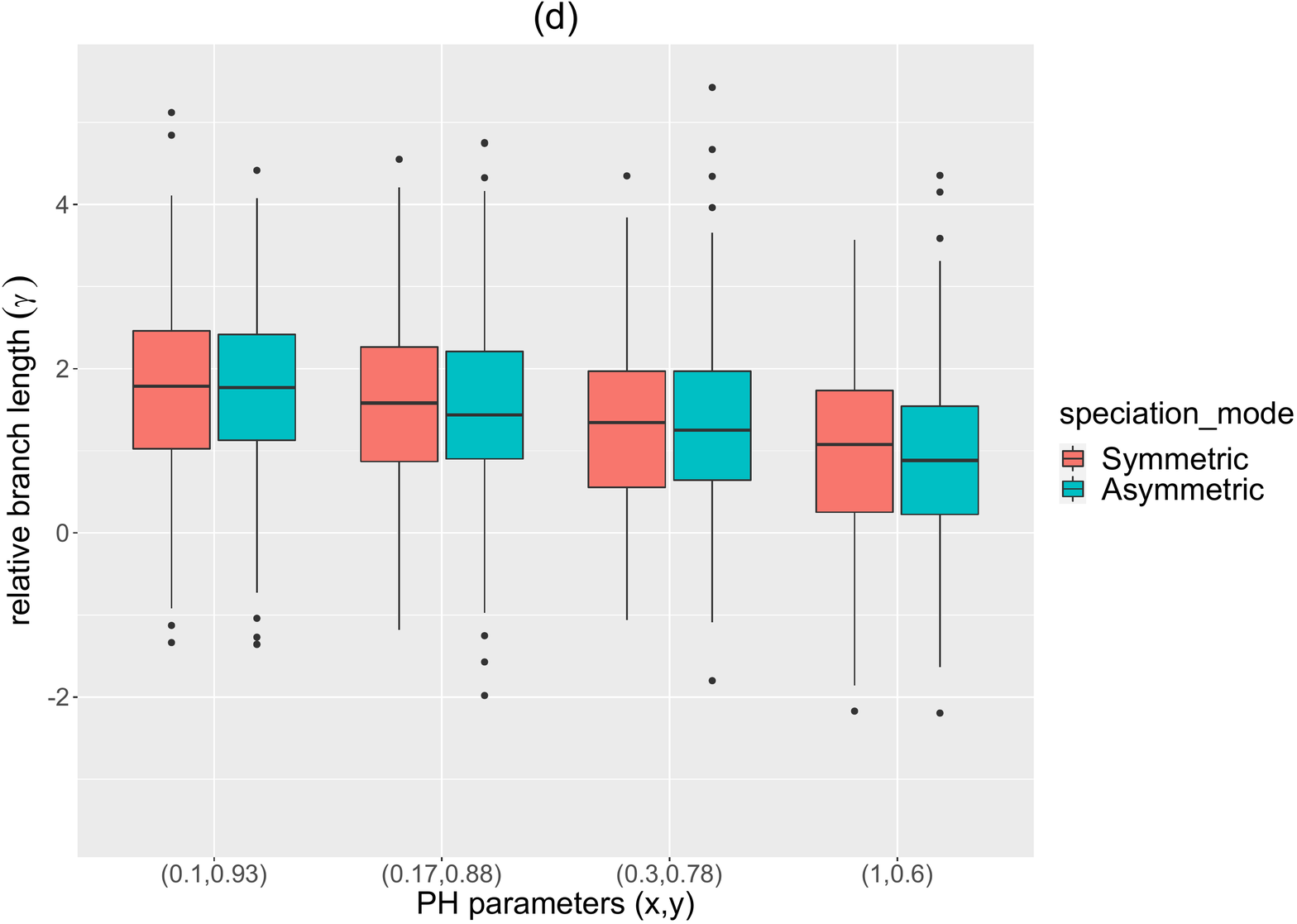}
		\end{subfigure}
		\caption{The effect of speciation and extinction processes on tree balance as measured by the $\beta$ statistic, and on branch lengths as measured by the $\gamma$ statistic. For each pair of parameters $(x,y)$ in Example~\ref{model} used to generate either times to speciation (in Fig.~\ref{Fig1}a,c) or times to extinction (in Fig.~\ref{Fig1}b,d), we simulated $300$ trees with $100$ extant tips. In Fig.~\ref{Fig1}a,c the parameters of the times to speciation are $(x,y) \in \{(0.1,0.93),(0.17,0.88),(0.3,0.78),(1,0.6)\}$ and mean speciation time is $E_{PH}(X)=2 $, while times to extinction are drawn from exponential distribution with some constant rate $\lambda$. In Fig.~\ref{Fig1}b,d times to speciation are drawn from exponential distribution with some constant rate $\lambda$, while the parameters of the times to extinctions are $(x,y) \in \{(0.1,0.93),(0.17,0.88),(0.3,0.78),(1,0.6)\}$ and mean extinction time is $E_{PH}(X)=2 $. The red dots show the $\beta$ statistic for sets of trees}
		\label{Fig1}
	\end{figure}

	We use the Coxian PH distribution in Example~\ref{model} for times to speciation or extinction events to study the effect of our model parameters on tree shapes. In Figure~\ref{Fig1}a,b, we show that tree balance, as measured by the median $\beta$ statistic, is affected by varying the parameters for times to speciation (Fig.~\ref{Fig1}a), while it is not significantly affected by the parameters for times to extinction (Fig.~\ref{Fig1}b). We also observe that changing the parameters of our macroevolutionary model leads to a wide range of tree balance, from unbalanced to balanced trees, as seen in Figure~\ref{Fig1}. Thus, it is possible to fit our model parameters to match the tree-shape statistics of empirical phylogenies.  
	
	We demonstrate the opposite effect of the model parameters on branch lengths, as measured by the $\gamma$ statistic, in Fig.~\ref{Fig1}c,d. Indeed, the $\gamma$ statistic are not affected by the parameters for times to speciation (Fig.~\ref{Fig1}c), while they are affected by the parameters for times to extinction (Fig.~\ref{Fig1}d).
	
	Finally, we note that we have not observed a significant difference in our results between the symmetric and asymmetric speciation modes~\citep{stadler2013recovering}. The choice between the two speciation modes did not affect tree balance and relative branch lengths.
	
	\subsection{The $\beta$ Statistic Computed Using Treesets Gives More Accurate Result}
	\label{sec:10}
	Below, we demonstrate that computing $\beta$ using treesets, as discussed earlier, is more accurate than computing $\beta$ for individual trees and then taking median value. To illustrate this, we consider a YH process to generate trees, and estimate $\beta$ using each of the two alternative methods, noting that for this process $\beta=0$ as shown in~\cite{aldous2001}.
	
\begin{figure}[!htbp]
		\centering     
		\begin{subfigure}[!htbp]{0.43\textwidth}
			\centering
			\hspace{-25mm}
			\includegraphics[width=8.4cm,height=23.4cm,keepaspectratio]{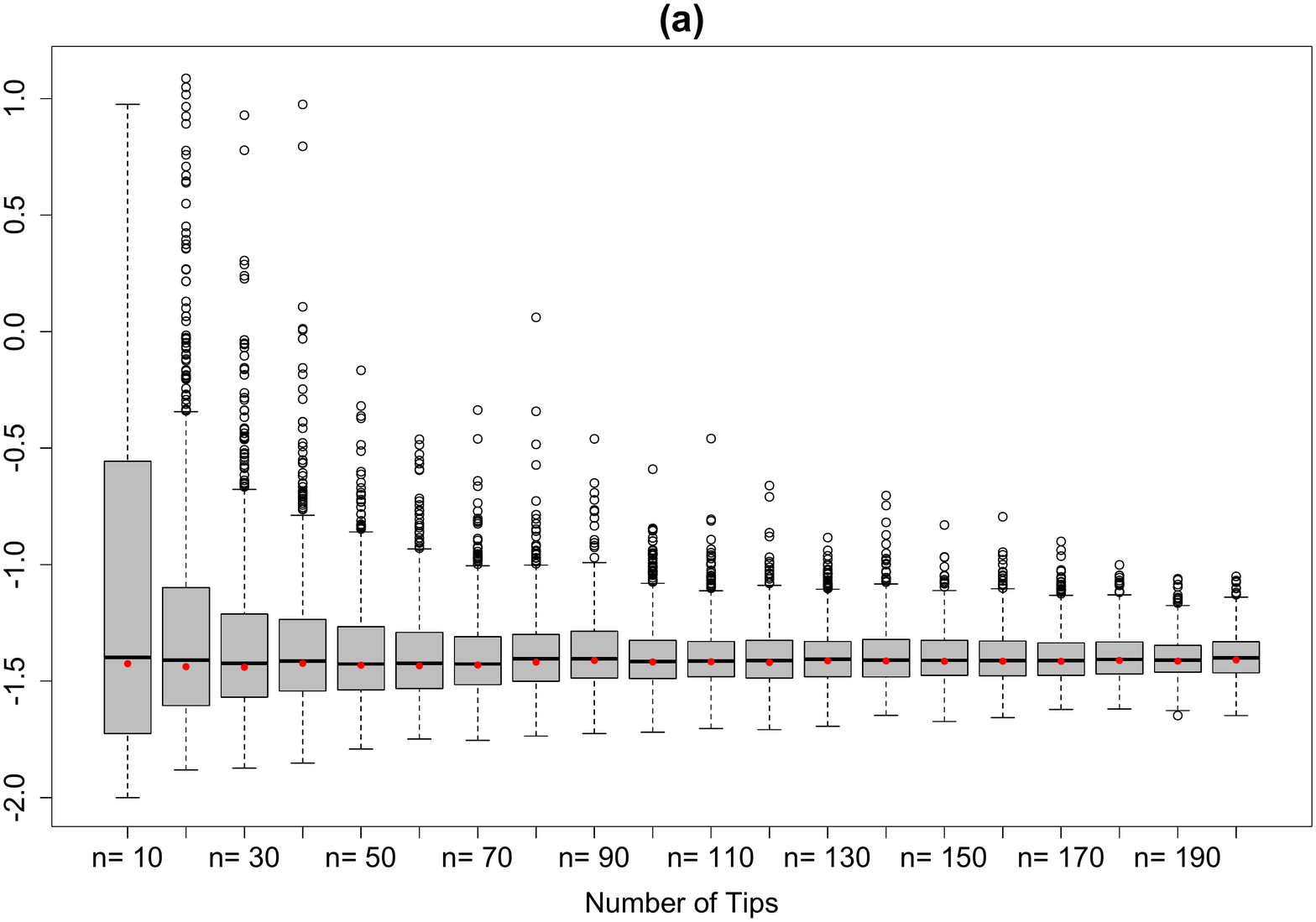}
		\end{subfigure}%
		~ 
		\begin{subfigure}[!htbp]{0.5\textwidth}
			\centering
			\hspace{25mm}
			\includegraphics[width=8.4cm,height=23.4cm,keepaspectratio]{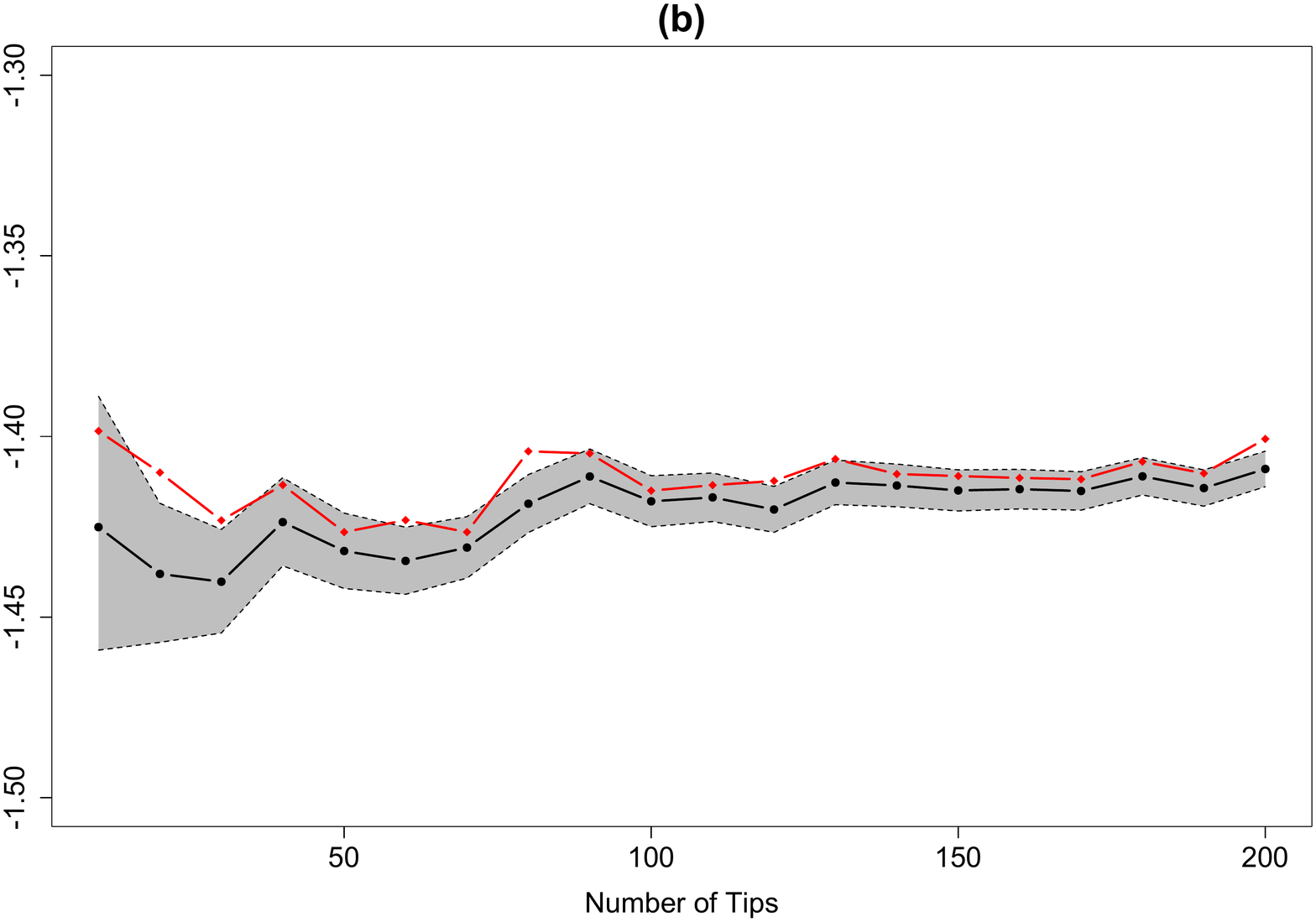}
		\end{subfigure}
		\par\bigskip
		~
		\begin{subfigure}[!htbp]{0.43\textwidth}
			\centering
			\hspace{-25mm}
			\includegraphics[width=8.4cm,height=23.4cm,keepaspectratio]{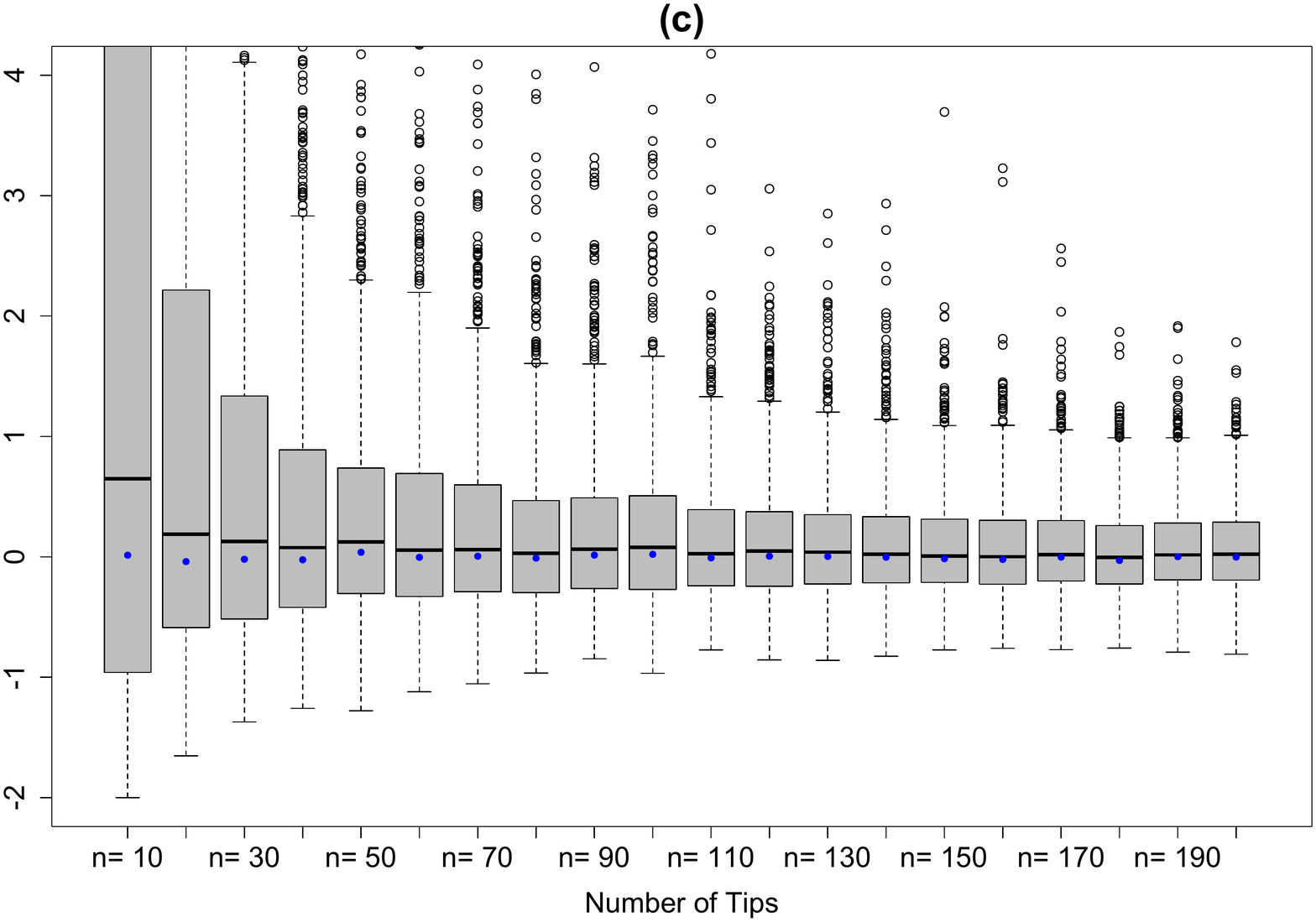}
		\end{subfigure}%
		~
		\begin{subfigure}[!htbp]{0.5\textwidth}
			\centering
			\hspace{25mm}
			\includegraphics[width=8.4cm,height=23.4cm,keepaspectratio]{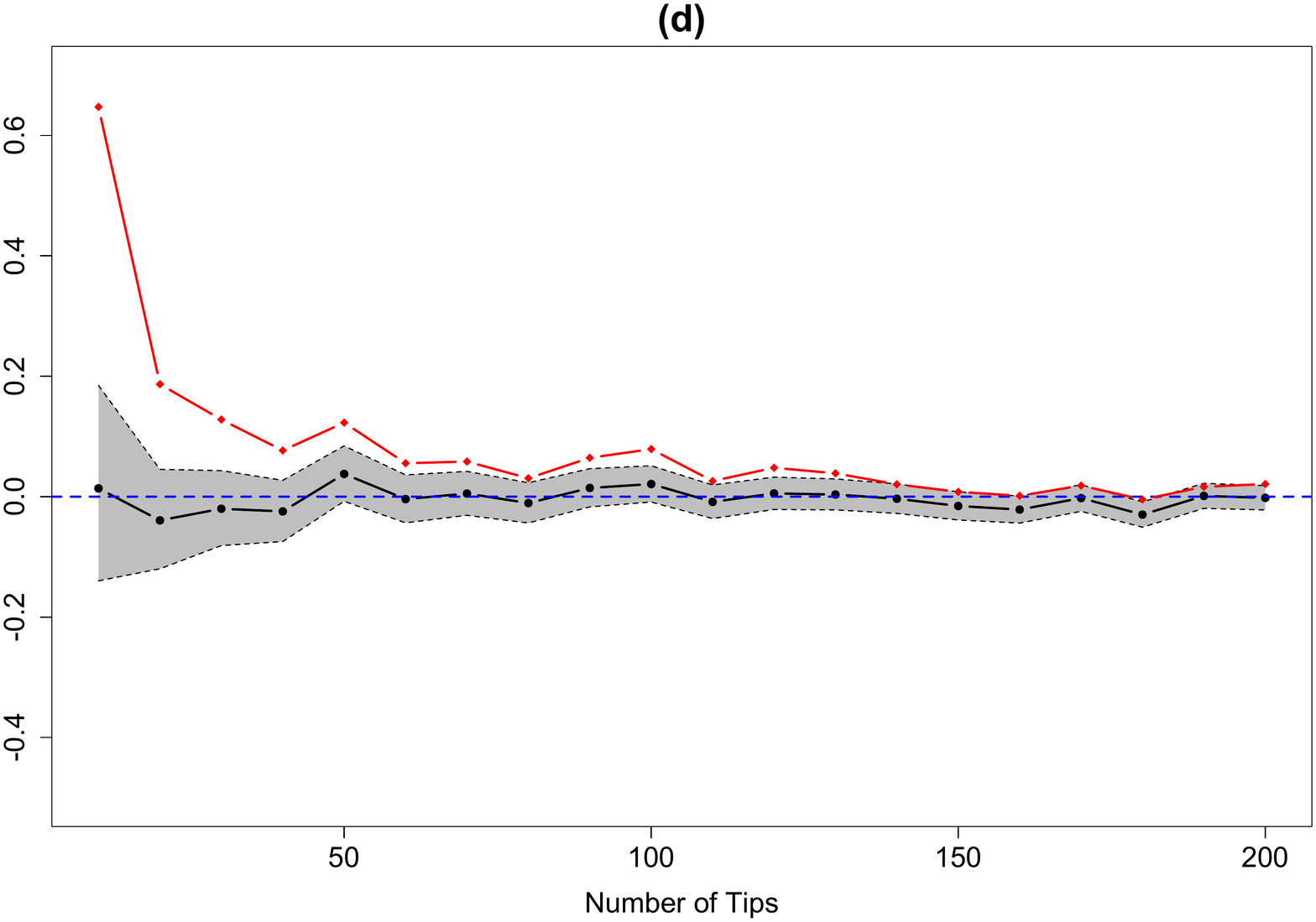}
		\end{subfigure}
		\captionsetup{justification=justified}
		\caption{Estimates of $\beta$ for individual trees with $n \in \{10,20,30,\cdots,200\}$ tips (Fig.~\ref{Fig3}a-d). Estimates of $\beta$ from treesets are indicated by red dots (Fig.~\ref{Fig3}a) and blue dots (Fig.~\ref{Fig3}c). Trees are simulated according to either Coxian PH distribution for times to speciation events (Fig.~\ref{Fig3}a) or the YH process (Fig.~\ref{Fig3}c). The area of $95\%$ confidence interval of $\beta$ values from treesets following Coxian PH distribution and the YH process are plotted in Fig.~\ref{Fig3}b,d respectively. The black lines represent the treeset $\beta$ values, and the grey area represents the confidence interval for each treeset $\beta$ value. The red lines represent the median $\beta$ values from individual trees. The blue-dashed line represents the theoretical $\beta$ value for the YH trees $\left(\beta=0\right)$}
		\label{Fig3}
	\end{figure}

	When estimating the value of $\beta$ for trees with small number of extant species, we obtained $\beta\approx 0$ when applying the first method (based on treesets), but $\beta>0$ when applying the second method. These results are summarised in Fig.~\ref{Fig3}. We conclude that the method based on treesets is more accurate, as evidenced by the $95\%$ confidence interval in Fig.~\ref{Fig3}d. We observe that $\beta$ values estimated from different sets of trees concentrate around $\beta = 0$, which agrees with the theoretical value for trees evolving under the YH model. A similar pattern is seen in Fig.~\ref{Fig3}b where the individual estimates of $\beta$ are biased upwards compared to the estimate based on sets of trees, particularly for trees with small numbers of tips.
	
	\subsection{Branch Lengths are Sufficient to Fit Parameters of Our Macroevolutionary Model}
	\label{sec:11}
\begin{figure}[!htbp]
		\centering     
		\captionsetup{width=\linewidth}
		\includegraphics[width=8.4cm,height=23.4cm,keepaspectratio]{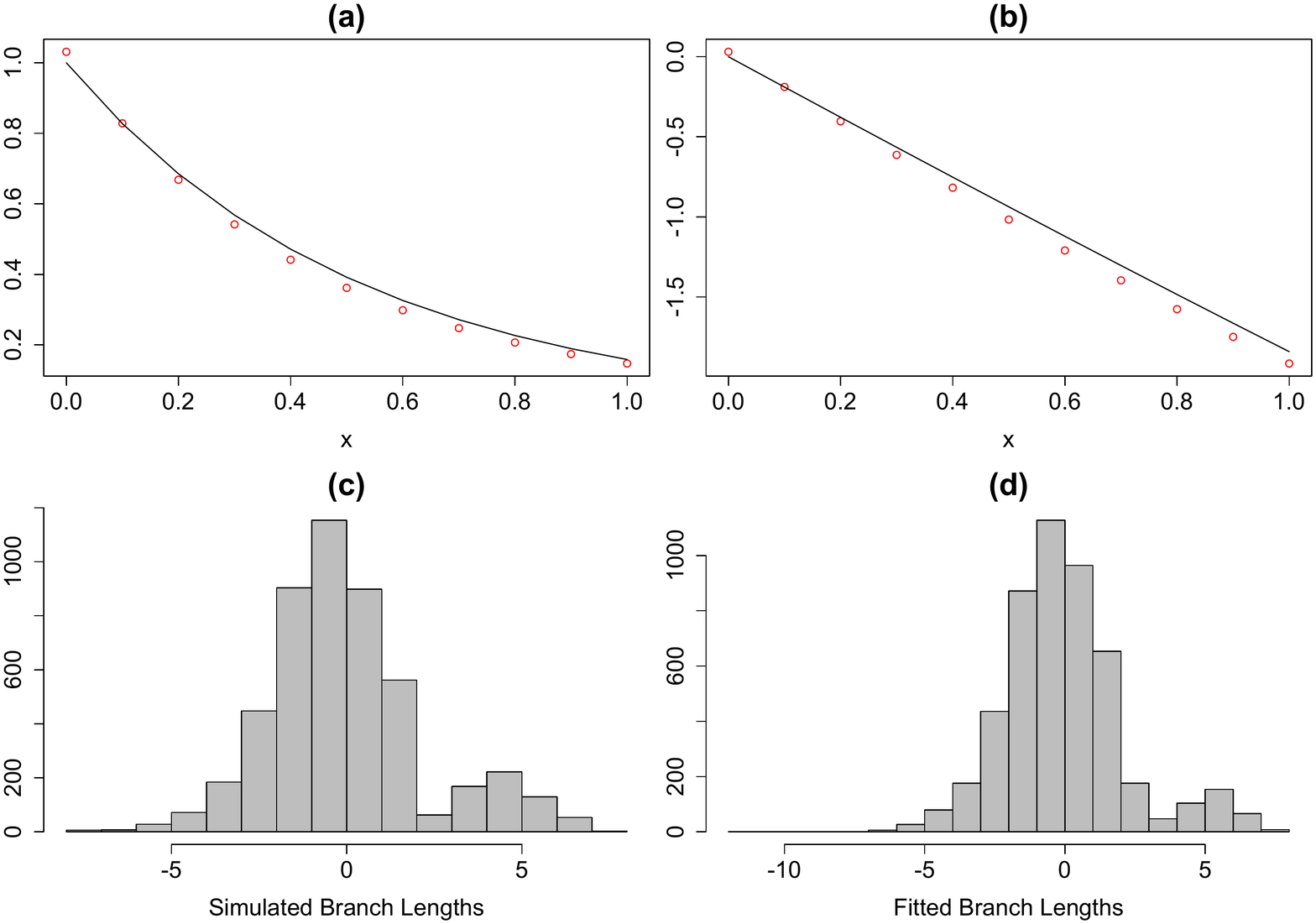}
		\caption{Macroevolutionary model is a good fit to simulated data. The density function of the PH distribution in Eq.~\ref{simplecox} (black line in Fig.~\ref{fit4states}a) and the values of the fitted density function in Example~\ref{model} (red dots in Fig.~\ref{fit4states}a). The corresponding log values are in Fig.~\ref{fit4states}b. The histograms of the simulated and fitted branch length distributions, shown in log scale, are displayed in Fig.~\ref{fit4states}c,d respectively. We used four non-absorbing states in each model}
		\label{fit4states}
	\end{figure}

	We found that branch lengths are sufficient to obtain a good fit of parameters of our macroevolutionary model to reconstructed trees. To study this, first we generated $50$ trees each with $50$ extant species for a range of distribution shapes using suitable choice of parameters of PH distribution  based on Eq.~\ref{simplecox}. Next, we applied Eq.~\ref{prob_treenoext}-\ref{prob_treeset} to fit the parameters of our model in Example~\ref{model} to this data, using only branch lengths. The results of this analysis are in Fig.~\ref{fit4states}. 
	
	We also found that a model with four non-absorbing states is a good fit, and that having more than four non-absorbing states does not significantly improve its fitness (See the appendix).
	
\begin{figure}[!htbp]
		\centering
		\includegraphics[width=8.4cm,height=23.4cm,keepaspectratio]{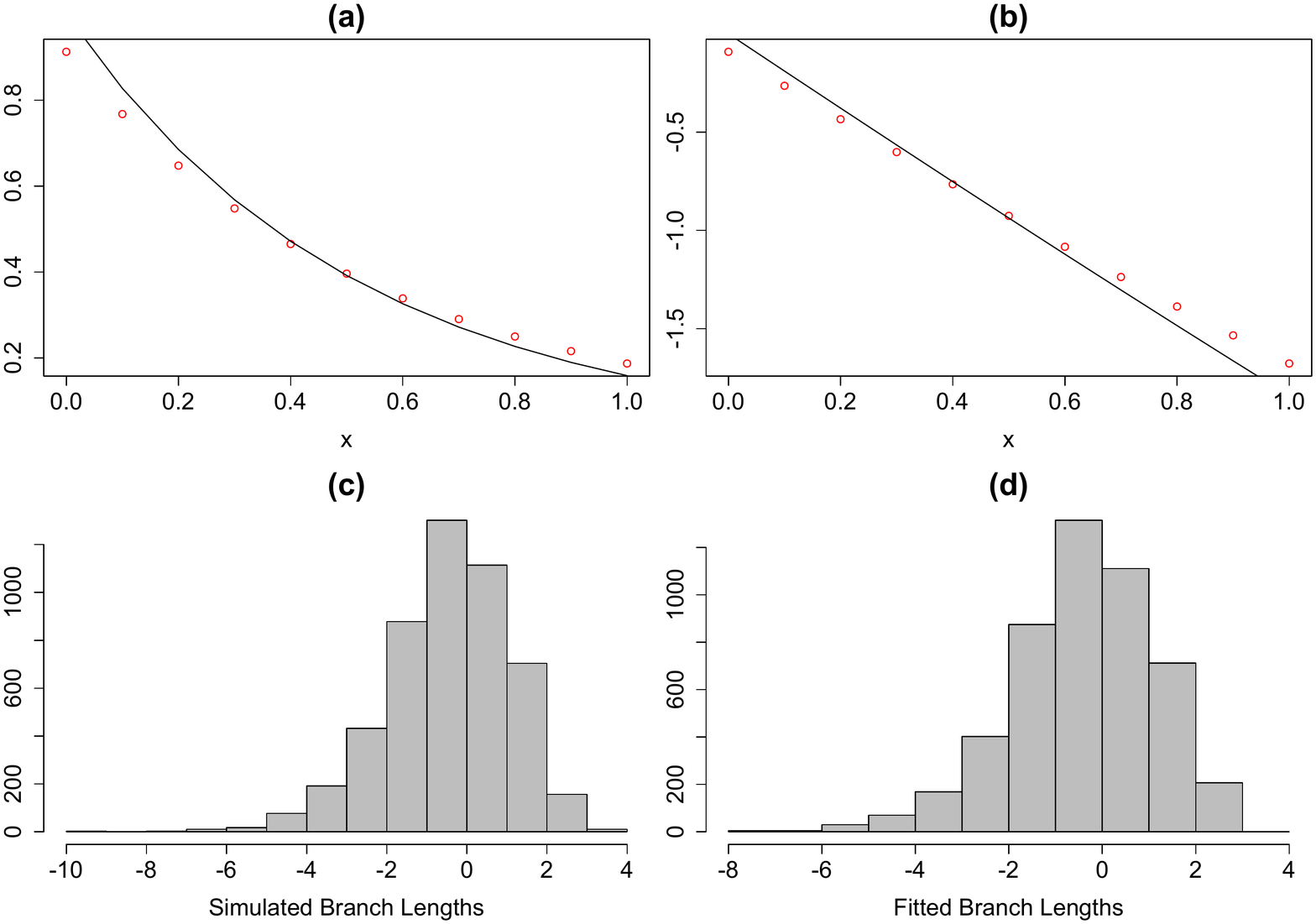}
		\caption{Including lengths of extinct branches produces biased estimates of the model parameters. Here, extinction events follow an exponential distribution with rate $\lambda=0.1$. The density function of the PH distribution in Eq.~\ref{simplecox} (black line in Fig.~\ref{Fig6}a) and the values of the fitted density function in Example~\ref{model} (red dots in Fig.~\ref{Fig6}a). The corresponding log values are in Fig.~\ref{Fig6}b. The histograms of the simulated and fitted branch length distributions, shown in log scale, are displayed in Fig.~\ref{Fig6}c,d respectively. We used four non-absorbing states in each model
		}
		\label{Fig6}
	\end{figure}

\begin{figure}[h!]
		\centering
		\captionsetup{width=\linewidth}
		\includegraphics[width=8.4cm,height=23.4cm,keepaspectratio]{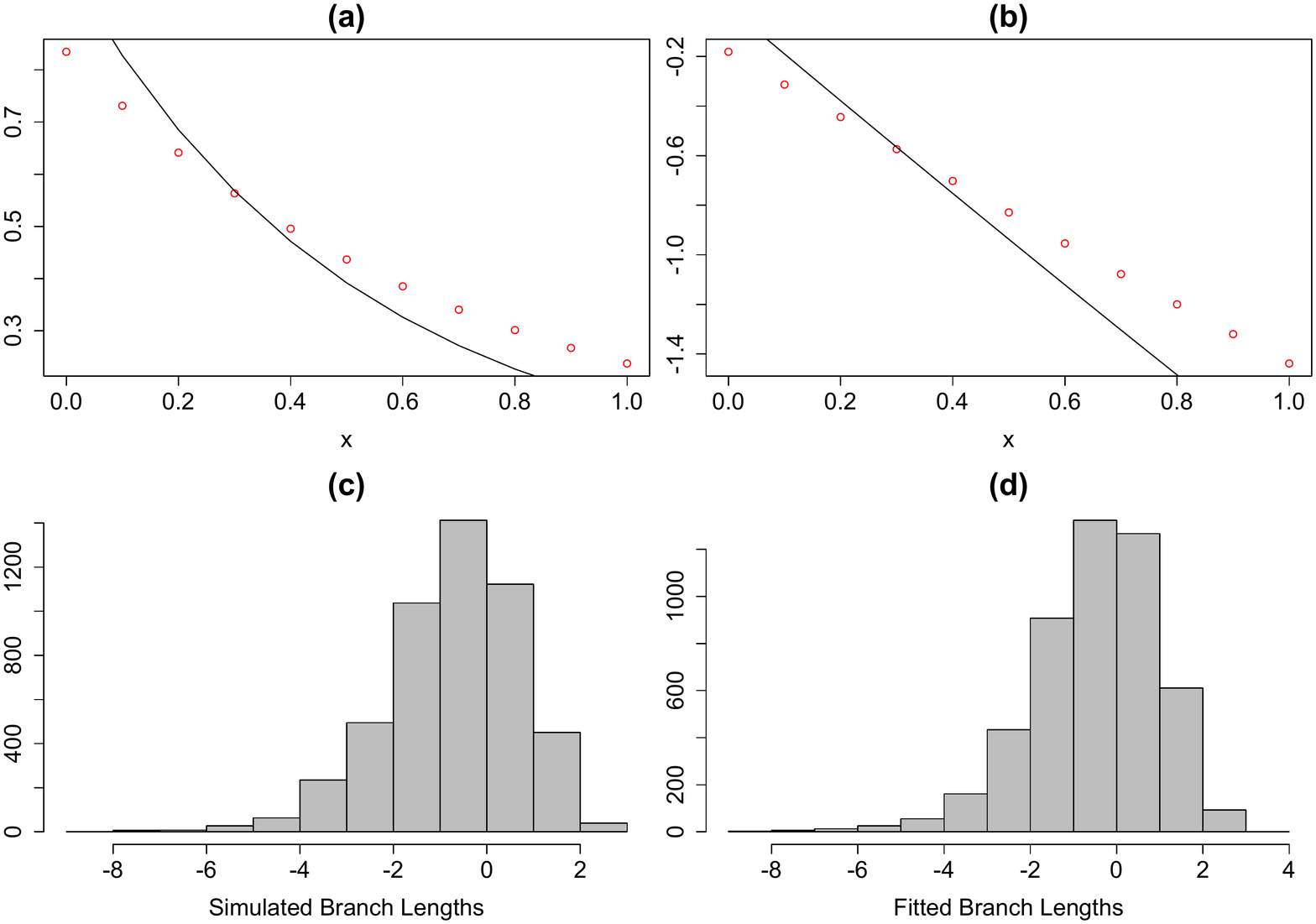}
		\caption{Increasing extinction rate $\lambda$ results in more biased estimates of the model parameters. Here, extinction events follow an exponential distribution with rate $\lambda=0.4$. Compare this with Fig.~\ref{Fig6} in which we had $\lambda=0.1$. The density function of the PH distribution in Eq.~\ref{simplecox} (black line in Fig.~\ref{Fig7}a) and the values of the fitted density function in Example~\ref{model} (red dots in Fig.~\ref{Fig7}a). The corresponding log values are in Fig.~\ref{Fig7}b. The histograms of the simulated and fitted branch length distributions, shown in log scale, are displayed in Fig.~\ref{Fig7}c,d respectively. We used four non-absorbing states in each model
		}
		\label{Fig7}
	\end{figure}

	We note that including branch lengths from reconstructed phylogenetic trees where extinction events are considered may produce biased estimates of the model parameters, since Eq.~\ref{prob_treenoext}-\ref{prob_treeset} are then insufficient to obtain a good fit. The bias becomes more apparent as we increase the extinction rate (Fig.~\ref{Fig6} and~\ref{Fig7}).
	
	\subsection{Fitting Model Parameters to Empirical Phylogenies}
	\label{sec:12}
	Here, we fit branch lengths from the whole reconstructed squamate phylogeny \citep{pyron2013phylogeny} and branch lengths from three different clades of the tree, namely the \textit{gekkota} clade ($1,318$ branches), the \textit{iguania} clade ($1,936$  branches), and the \textit{anguimorpha} clade ($200$ branches), to the following models. We consider three models where the speciation process follows PH distributions in Definition~\ref{generalCox} and Examples~\ref{model}--\ref{model2}, respectively, and the other two models where the speciation process follows an exponential distribution and a Weibull distribution, respectively. The results are summarised in Table~\ref{Fig8} below. 
	
	As demonstrated in Table~\ref{Fig8} below, the model in which speciation process follows a general Coxian PH distribution in Definition~\ref{generalCox}, provides the best fit to the {\em majority} of the cases, compared to the other distributions.
	
	To see how each model performs in a larger tree, we also fit branch lengths from four different clades of the reconstructed angiosperm phylogeny \citep{zanne2014three}. The four different clades we use are: the \textit{monocotyledoneae} clade ($14,118$ branches), the \textit{magnoliidae} clade ($2,092$ branches), the \textit{superrosidae} clade ($11,323$ branches), and the \textit{superasteridae} clade ($20,016$ branches). The results are summarised in Table~\ref{Fig9} below. 
	
	\begin{table}
		\caption{Model selection is based on the likelihood of observing a tree with no extinction process, as defined in Eq.~\ref{prob_treenoext}. We select model that has the lowest AIC value as the base model and compute $\Delta$AIC$=\text{AIC}_{\text{other model}}-\text{AIC}_{\text{best model}}$. We use branch lengths from (a) the whole reconstructed squamate tree; and from different clades from the tree, namely (b) the \textit{gekkota} clade, (c) the \textit{iguania} clade, and (d) the \textit{anguimorpha} clade. Model with a Coxian PH distribution in Definition~\ref{generalCox} provides best fit in majority of the cases. }
		\label{Fig8}
		\begin{subtable}[!htbp]{0.8\textwidth}
			\centering
			\captionsetup{width=\linewidth}
			\caption*{(a)}
			\hspace*{0cm} \begin{tabular}{|l|c|c|c|c|c|}
				\hline
				Model & \# branches & \# parameters & LogL & AIC &$\Delta$AIC\\
				\hline
				General Coxian PH distribution & \multirow{5}*{$8322$} 
				& $7$ & $6503.684$ & $-12993.367$ & $0$\\
				\cline{3-6}
				PH model assuming decreasing rate & 	   & $3$ & $6466.357$ & $-12926.713$ & $66.654$\\
				\cline{3-6}
				PH model assuming increasing rate & 	   & $3$ & $2026.811$ & $-4047.623$ & $8945.744$\\
				\cline{3-6}
				Exponential distribution & 				   & $1$ & $6032.986$ & $-12063.972$ & $929.395$\\
				\cline{3-6}
				Weibull distribution & 	                   & $2$ & $6213.098$ & $-12422.196$ & $571.171$\\
				\hline
			\end{tabular}
			
		\end{subtable}
		
		\bigskip
		\begin{subtable}[!htbp]{0.8\textwidth}
			\centering
			\caption*{(b)}
			\hspace*{0cm} \begin{tabular}{|l|c|c|c|c|c|}
				\hline
				Model & \# branches & \# parameters & LogL & AIC &$\Delta$AIC\\
				\hline
				General Coxian PH distribution & \multirow{5}*{$1318$} 
				& $7$ & $718.5753$ & $-1423.1506$ & $0$\\
				\cline{3-6}
				PH model assuming decreasing rate & 	   & $3$ & $714.1257$ & $-1422.2514$ & $0.8992$\\
				\cline{3-6}
				PH model assuming increasing rate & 	   & $3$ & $255.9474$ & $-505.8947$ & $917.2559$\\
				\cline{3-6}
				Exponential distribution & 				   & $1$ & $648$ & $-1294$ & $129.1506$\\
				\cline{3-6}
				Weibull distribution & 	                   & $2$ & $668.9134$ & $-1333.8268$ & $89.3238$\\
				\hline
			\end{tabular}
		\end{subtable}
		
		\bigskip
		\begin{subtable}[!htbp]{0.8\textwidth}
			\centering
			\caption*{(c)}
			\hspace*{0cm} \begin{tabular}{|l|c|c|c|c|c|}
				\hline
				Model & \# branches & \# parameters & LogL & AIC &$\Delta$AIC\\
				\hline
				General Coxian PH distribution & \multirow{5}*{$1936$} 
				& $7$ & $1598.5608$ & $-3183.1216$ & $0$\\
				\cline{3-6}
				PH model assuming decreasing rate & 	   & $3$ & $1587.7247$ & $-3169.4495$ & $13.6721$\\
				\cline{3-6}
				PH model assuming increasing rate & 	   & $3$ & $489.5528$ & $-973.1056$ & $2210.016$\\
				\cline{3-6}
				Exponential distribution & 				   & $1$ & $1520.4084$ & $-3038.8167$ & $144.3049$\\
				\cline{3-6}
				Weibull distribution & 	                   & $2$ & $1538.4562$ & $-3072.9125$ & $110.2091$\\
				\hline
			\end{tabular}
		\end{subtable}
		
		\bigskip
		\begin{subtable}[!htbp]{0.8\textwidth}
			\centering
			\caption*{(d)}
			\hspace*{0cm} \begin{tabular}{|l|c|c|c|c|c|}
				\hline
				Model & \# branches & \# parameters & LogL & AIC &$\Delta$AIC\\
				\hline
				General Coxian PH distribution & \multirow{5}*{$200$} 
				& $7$ & $153.2953$ & $-292.5907$ & $2.8832$\\
				\cline{3-6}
				PH model assuming decreasing rate & 	   & $3$ & $148.8397$ & $-291.6794$ & $3.7945$\\
				\cline{3-6}
				PH model assuming increasing rate & 	   & $3$ & $48.7780$ & $-91.5561$ & $203.9178$\\
				\cline{3-6}
				Exponential distribution & 				   & $1$ & $146.9074$ & $-291.8148$ & $3.6591$\\
				\cline{3-6}
				Weibull distribution & 	                   & $2$ & $149.7369$ & $-295.4739$ & $0$\\
				\hline
			\end{tabular}
		\end{subtable}
	\end{table}
	
	\begin{table}
		\caption{Model with a Coxian PH distribution in Definition~\ref{generalCox} provides best fit in all of the cases. Model selection is based on the likelihood of observing a tree with no extinction process, as defined in Eq.~\ref{prob_treenoext}. We select model that has the lowest AIC value as the base model and compute $\Delta$AIC$=\text{AIC}_{\text{other model}}-\text{AIC}_{\text{best model}}$. We use branch lengths from different clades of the reconstructed angiosperm phylogeny, namely (a) the \textit{monocotyledoneae} clade, (b) \textit{magnoliidae} clade, (c) \textit{superasteridae} clade and (d) the \textit{superrosidae} clade}
		\label{Fig9}
		\begin{subtable}[!htbp]{0.8\textwidth}
			\centering
			\caption*{(a)}
			\hspace*{0cm} \begin{tabular}{|l|c|c|c|c|c|}
				\hline
				Model & \# branches & \# parameters & LogL & AIC &$\Delta$AIC\\
				\hline
				General Coxian PH distribution & \multirow{5}*{$14118$} 
				& $7$ & $-18439.46$ & $36892.91$ & $0$\\
				\cline{3-6}
				PH model assuming decreasing rate & 	   & $3$ & $-18788.38$ & $37582.76$ & $689.85$\\
				\cline{3-6}
				PH model assuming increasing rate & 	   & $3$ & $-45591.17$ & $91188.34$ & $54295.43$\\
				\cline{3-6}
				Exponential distribution & 				   & $1$ & $-22149.68$ & $44301.37$ & $7408.46$\\
				\cline{3-6}
				Weibull distribution & 	                   & $2$ & $-18633.07$ & $37270.14$ & $377.23$\\
				\hline
			\end{tabular}
			
		\end{subtable}
		
		\bigskip
		\begin{subtable}[!htbp]{0.8\textwidth}
			\centering
			\caption*{(b)}
			\hspace*{0cm} \begin{tabular}{|l|c|c|c|c|c|}
				\hline
				Model & \# branches & \# parameters & LogL & AIC &$\Delta$AIC\\
				\hline
				General Coxian PH distribution & \multirow{5}*{$2092$} 
				& $7$ & $-3347.301$ & $6708.602$ & $0$\\
				\cline{3-6}
				PH model assuming decreasing rate & 	   & $3$ & $-3476.505$ & $6959.010$ & $250.408$\\
				\cline{3-6}
				PH model assuming increasing rate & 	   & $3$ & $-8925.996$ & $17857.993$ & $11149.391$\\
				\cline{3-6}
				Exponential distribution & 				   & $1$ & $-3633.437$ & $7268.873$ & $560.271$\\
				\cline{3-6}
				Weibull distribution & 	                   & $2$ & $-3395.168$ & $6794.337$ & $85.735$\\
				\hline
			\end{tabular}
		\end{subtable}
		
		\bigskip
		\begin{subtable}[!htbp]{0.8\textwidth}
			\centering
			\caption*{(c)}
			\hspace*{0cm} \begin{tabular}{|l|c|c|c|c|c|}
				\hline
				Model & \# branches & \# parameters & LogL & AIC &$\Delta$AIC\\
				\hline
				General Coxian PH distribution & \multirow{5}*{$20016$} 
				& $7$ & $-29747.59$ & $59509.18$ & $0$\\
				\cline{3-6}
				PH model assuming decreasing rate & 	   & $3$ & $-30533.85$ & $61073.71$ & $1564.53$\\
				\cline{3-6}
				PH model assuming increasing rate & 	   & $3$ & $-59551.17$ & $119108.33$ & $59599.15$\\
				\cline{3-6}
				Exponential distribution & 				   & $1$ & $-33668.54$ & $67339.07$ & $7829.89$\\
				\cline{3-6}
				Weibull distribution & 	                   & $2$ & $-30064.20$ & $60132.40$ & $623.22$\\
				\hline
			\end{tabular}
		\end{subtable}
		
		\bigskip
		\begin{subtable}[!htbp]{0.8\textwidth}
			\centering
			\caption*{(d)}
			\hspace*{0cm} \begin{tabular}{|l|c|c|c|c|c|}
				\hline
				Model & \# branches & \# parameters & LogL & AIC &$\Delta$AIC\\
				\hline
				General Coxian PH distribution & \multirow{5}*{$11323$} 
				& $7$ & $-29977.32$ & $59968.64$ & $0$\\
				\cline{3-6}
				PH model assuming decreasing rate & 	   & $3$ & $-30717.66$ & $61441.32$ & $1472.68$\\
				\cline{3-6}
				PH model assuming increasing rate & 	   & $3$ & $-72613.88$ & $145233.76$ & $85265.12$\\
				\cline{3-6}
				Exponential distribution & 				   & $1$ & $-33136.25$ & $66274.49$ & $6305.85$\\
				\cline{3-6}
				Weibull distribution & 	                   & $2$ & $-30183.83$ & $60371.67$ & $403.03$\\
				\hline
			\end{tabular}
		\end{subtable}
	\end{table}
	
	As demonstrated in the Table~\ref{Fig9}, the model in which speciation process follows a general Coxian distribution in Definition~\ref{generalCox}, provides the best fit to {\em all} of the cases, compared to the other distributions, including models with exponential and Weibull distributions. 
	
\begin{figure}[!htbp]
		\centering
		\captionsetup{width=\linewidth}     
		\includegraphics[width=8.4cm,height=23.4cm,keepaspectratio]{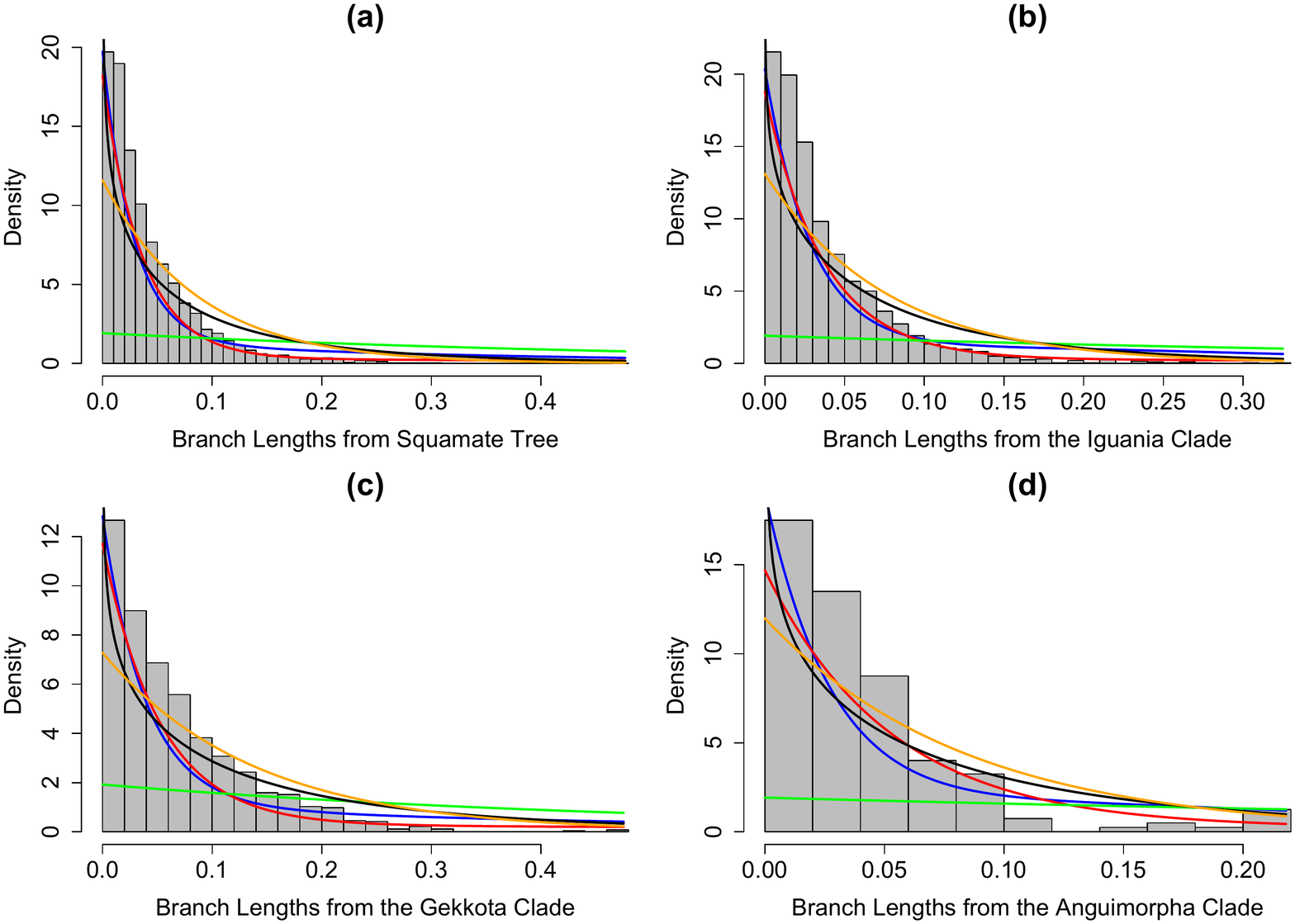}
		\caption{Histograms of empirical branch length density from the whole squamate tree (Fig.~\ref{Fig10}a), the \textit{iguania} clade (Fig.~\ref{Fig10}b), the \textit{gekkota} clade (Fig.~\ref{Fig10}c), and the \textit{anguimorpha} clade (Fig.~\ref{Fig10}d) with the fitted branch length densities from the five distributions mentioned above. The blue line is the fitted density using the general Coxian PH distribution defined in Definition~\ref{generalCox}, the red line is the fitted density using an example of a Coxian PH distribution defined in Example~\ref{model}, the green line is the fitted density using an example of a Coxian PH distribution defined in Example~\ref{model2}, and black and orange lines are the fitted density using Weibull and exponential distributions, respectively}
		\label{Fig10}
	\end{figure}

\begin{figure}[!htbp]
		\centering
		\captionsetup{width=\linewidth}     
		\includegraphics[width=8.4cm,height=23.4cm,keepaspectratio]{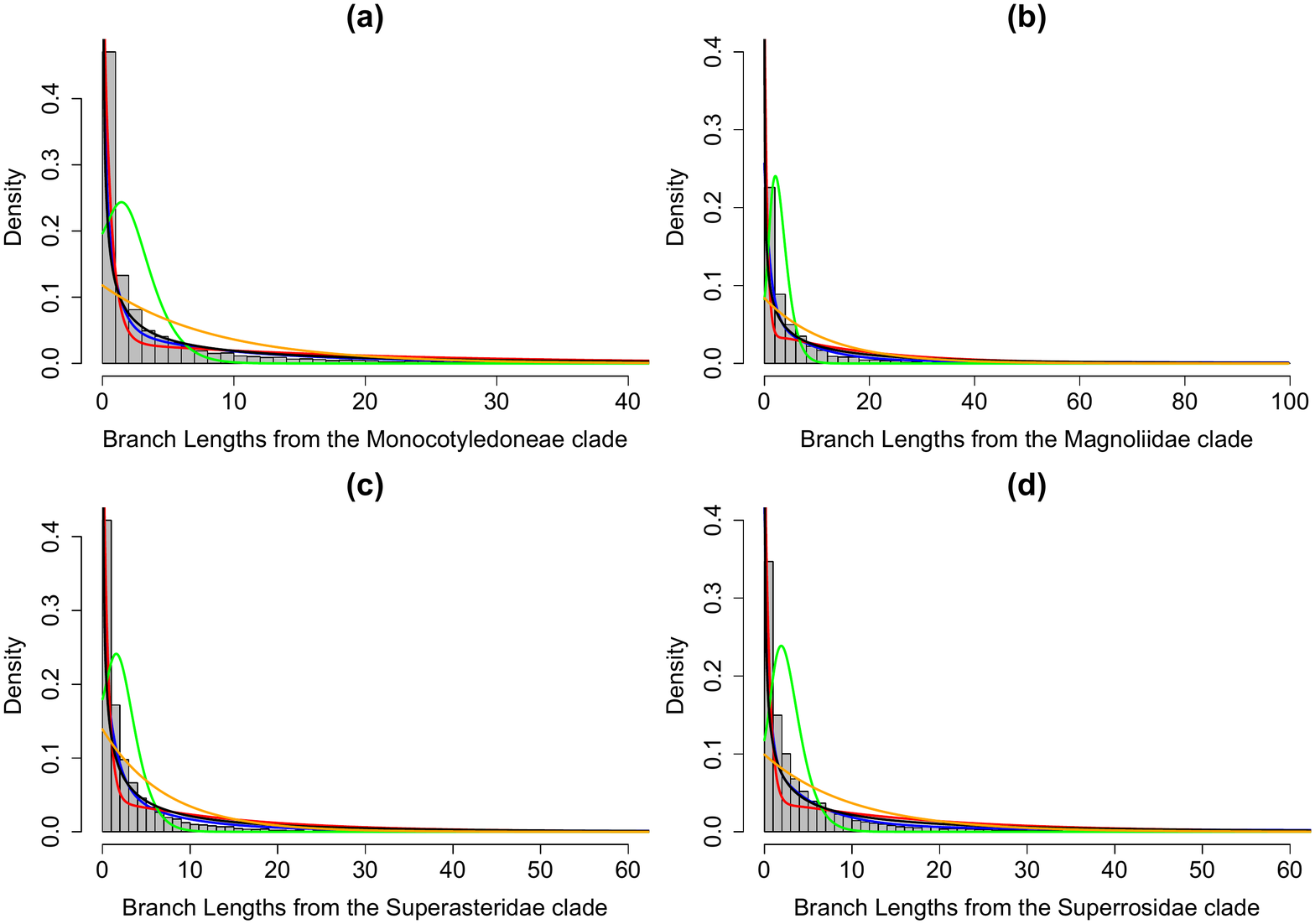}
		\caption{Histograms of empirical branch length density from the \textit{monocotyledoneae} clade (Fig.~\ref{Fig11}a), the \textit{magnoliidae} (Fig.~\ref{Fig11}b), the \textit{superasteridae} clade (Fig.~\ref{Fig11}c) and the \textit{superrosidae} clade (Fig.~\ref{Fig11}d) with the fitted branch length densities from the five distributions mentioned earlier. The blue line is the fitted density using a general Coxian PH distribution defined in Definition~\ref{generalCox}, the red line is the fitted density using an example of a Coxian PH distribution defined in Example~\ref{model}, the green line is the fitted density using an example of a Coxian PH distribution defined in Example~\ref{model2}, and black and orange lines are the fitted density using Weibull and exponential distributions, respectively}
		\label{Fig11}
	\end{figure}

	Next, we perform absolute goodness-of-fit test to see how each model differs from the empirical data by plotting probability density functions of branch lengths and empirical branch lengths, respectively, from the two phylogenies. The results are summarised in Fig.~\ref{Fig10}--\ref{Fig11}. 
	
	We observe that a general Coxian PH distribution in Definition~\ref{generalCox} generally fits better compared to the other distributions, including Weibull and exponential distributions (Fig.~\ref{Fig10} and Fig.~\ref{Fig11}). However, at the start of the histograms, none of the distributions fit well. We hypothesize that this result could due to ignoring extinction events in the models. 
	
	We also observe that the density of fitted distribution of Example~\ref{model2} does not follow the shape of the empirical histograms for all clades in the squamate as well as angiosperm trees. Since the structure of this distribution is constructed so that it corresponds to increasing speciation rates, this result suggests that speciation rates tend to decrease over time, in both phylogenies.
	
\begin{figure}[!htbp]
		\centering     
		\captionsetup{width=\linewidth} 
		\includegraphics[width=8.4cm,height=23.4cm,keepaspectratio]{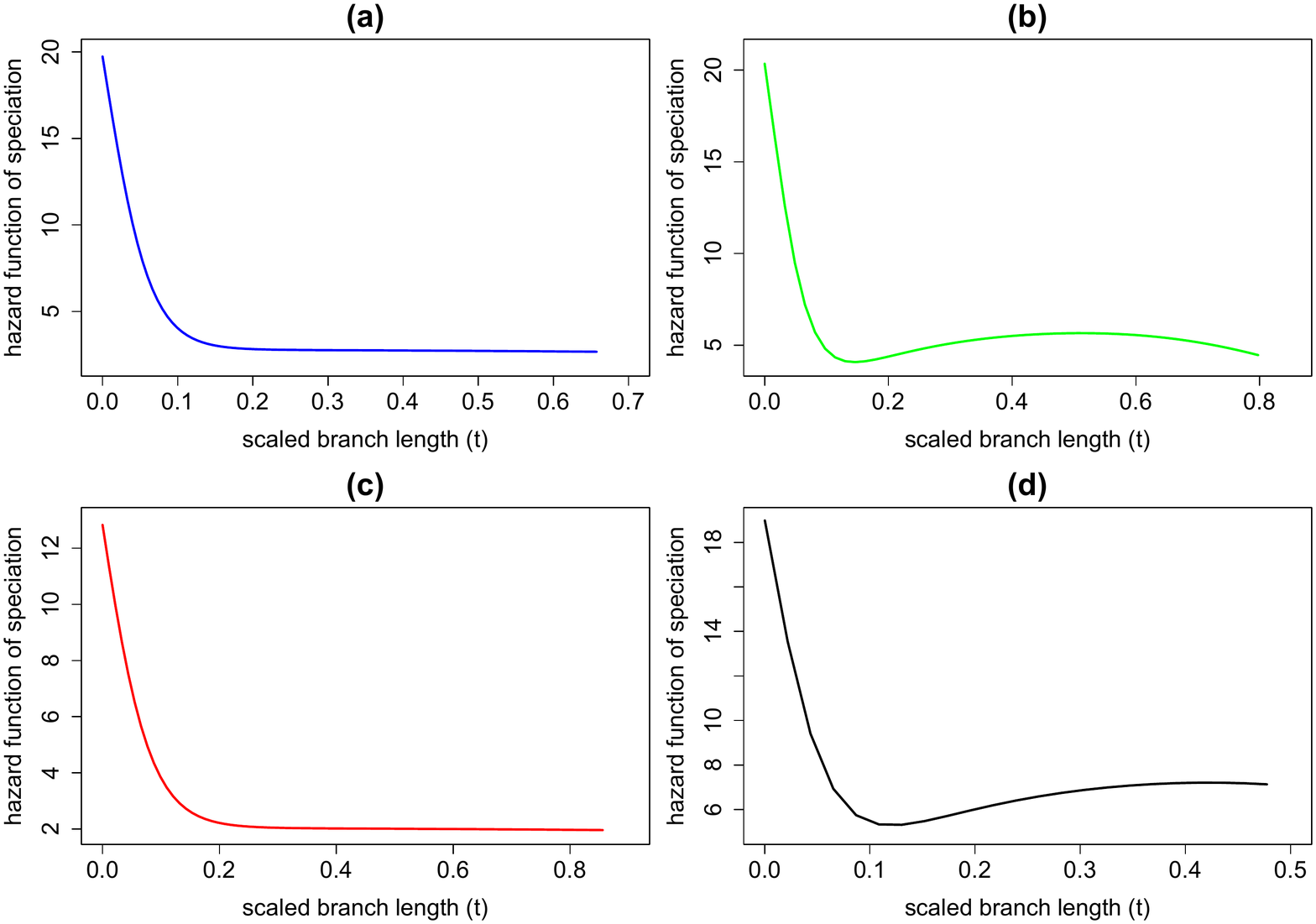}
		\caption{Hazard plots for speciation events using the best-fitting general Coxian PH distribution for the whole squamate tree (Fig.~\ref{Fig12}a), the \textit{iguania} clade (Fig.~\ref{Fig12}b), the \textit{gekkota} clade (Fig.~\ref{Fig12}c), and the \textit{anguimorpha} clade (Fig.~\ref{Fig12}d). For each tree, branches are scaled by dividing each branch length leading to speciation event with height of the tree}
		\label{Fig12}
	\end{figure}

\begin{figure}[!htbp]
		\centering     
		\captionsetup{width=\linewidth} 
		\includegraphics[width=8.4cm,height=23.4cm,keepaspectratio]{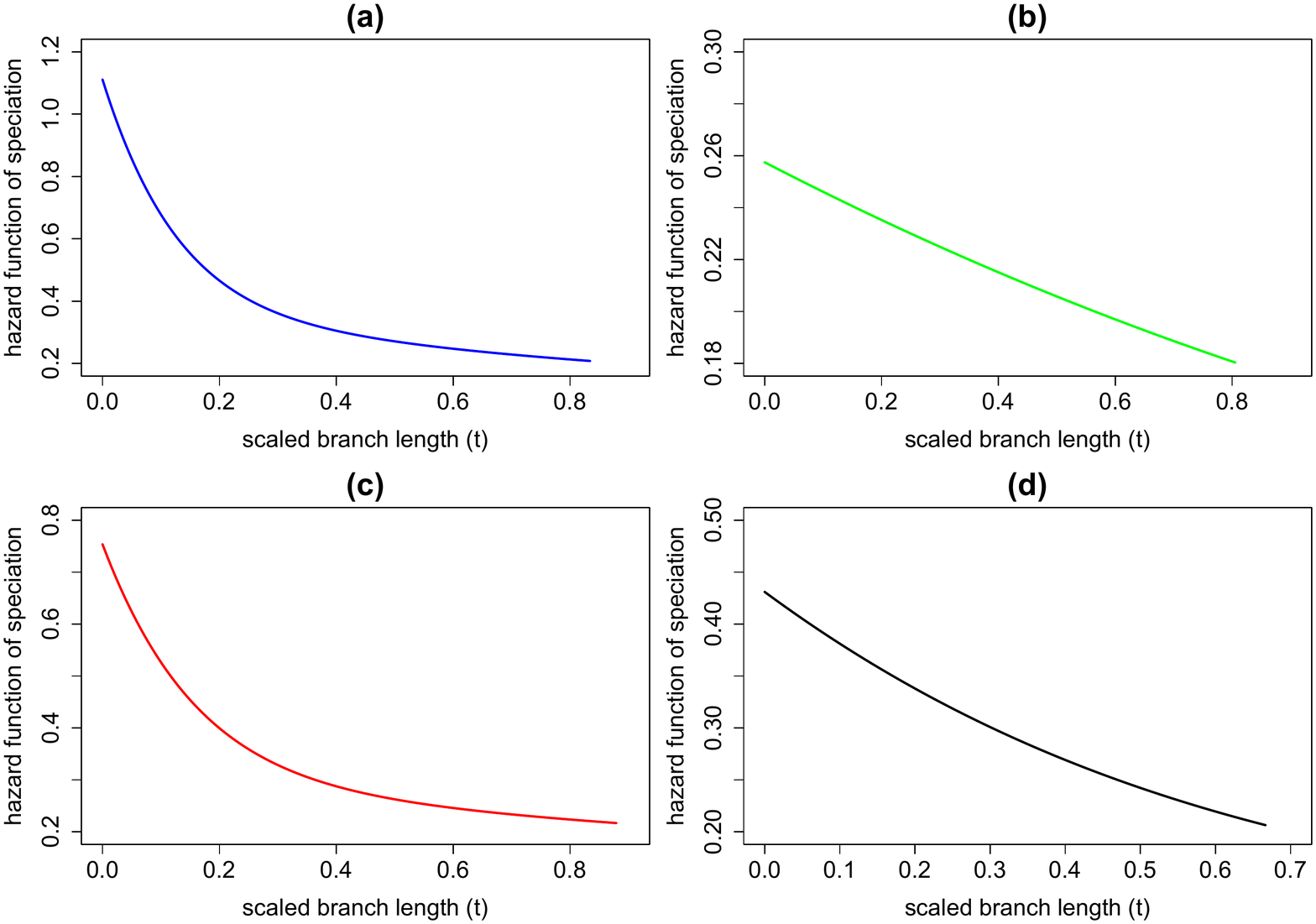}
		\caption{Hazard plots for speciation events using the best-fitting general Coxian PH distribution for the \textit{monocotyledoneae} clade (Fig.~\ref{Fig13}a), the \textit{magnoliidae} clade (Fig.~\ref{Fig13}b), the \textit{superasteridae} clade (Fig.~\ref{Fig13}c), and \textit{superrosidae} clade (Fig.~\ref{Fig13}d)}
		\label{Fig13}
	\end{figure}

	We note that in most clades, speciation events tend to occur more frequently at shorter branches, as demonstrated by the hazard plots for speciation from the best-fitting general Coxian PH distribution for each said clade and phylogenies  (Fig.~\ref{Fig12}a,c and \ref{Fig13}). This result implies high speciation rates on shorter branches of those clades. The exceptions to this are the \textit{iguania} and the \textit{anguimorpha} clades, for which the speciation rates tend to decrease and then increase before tailing off. This result suggests that the species that just appears is more likely to speciate, although at some longer branches, we observe more speciation events (Fig.~\ref{Fig12}b,d).
	
	\newpage
	\section{Discussion and Conclusion}
	\label{sec:13}
	Our macroevolutionary model for phylogenetic trees where times to speciation or extinction events are drawn from a Coxian PH distribution can produce phylogenetic trees with a range of tree shapes. The model provides a good fit to empirical data compared to exponential and Weibull distributions. The idea of applying PH distributions is motivated by the following two properties. First, it is well known that PH distributions are dense in the field of all positive-valued distributions \citep{asmussen1996}, and thus they are very flexible when fitting to empirical distributions. Second, evolution of species tree can be modelled as a forward-in-time process which follows acyclic PH distribution. Thus, it is reasonable to use Coxian PH distributions, due to the fact that any acyclic PH distribution can be represented as a Coxian PH distribution \citep{cumani1982canonical,asmussen1996}.
	
	We have demonstrated that trees generated under our model have a wide range of balance as measured by the $\beta$ statistic (Fig.~\ref{Fig1}). Thus, it is possible to fit parameters of our model to empirical tree shapes. 
	
	Moreover, we observe from Fig.~\ref{Fig1} that the simulated trees tend to be unbalanced in shape, suggesting that trees generated using PH distributions are not uniform on ranked tree shapes (URT) trees. Therefore, our model is in the model class 4 identified in~\cite{lambert2013birth}, in which the speciation process depends on non-heritable trait (species age) and the extinction process follows an exponential distribution with some constant rate. 
	
	We have also proposed a method of computing the $\beta$ statistic based on sets of trees, in which each tree has the same number of extant species. We have demonstrated that computing the $\beta$ statistic based on individual trees can produce bias, particularly for trees with smaller numbers of extant species. In contract, computing the $\beta$ statistic based on our method, gives a more accurate result (Fig.~\ref{Fig3}). 
	
	In our simulations, we have found that tree balance is mainly controlled by the speciation process, and is largely invariant to the extinction process. In contrast, the relative branch lengths, as measured by the $\gamma$ statistic, are to a large extent controlled by the extinction process, but relatively invariant to the speciation process. We have also found that symmetric and asymmetric speciation modes have a minor effect on the tree shapes. These findings agree with the results in~\cite{Hagen2015} in which speciation and extinction processes were modelled using Weibull distribution.
	
	Furthermore, we have derived a likelihood expression for the probability of observing any reconstructed tree (Eq.~\ref{prob_treenoext}), and applied it to fitting models to simulated and empirical data, by applying maximum likelihood method. We have demonstrated that it is sufficient to use branch lengths to get best-fitting parameters (Fig.~\ref{fit4states}), and that four non-absorbing states in a PH distribution are also sufficient, since adding more non-absorbing states does not significantly impact the fitting (see the appendix for Fig.~$4$). 
	
	We note that fitting parameters based on branch lengths taken from trees that include extinction, produces some bias (Fig.~\ref{Fig6}). The bias becomes more apparent with increasing rates of extinction (Fig.~\ref{Fig7}). In future work, we will generalise Eq.~\ref{prob_treenoext} to include extinction. Once we derive a generalised likelihood function, we will compare its performance with likelihood functions that consider both speciation and extinction events, such as in \cite{rabosky2006likelihood}.
	
	Finally, we have fitted the parameters of our model to the empirical data consisting of branch lengths from various clades in the squamate and angiosperm reconstructed phylogenies \citep{pyron2013phylogeny,zanne2014three}. Importantly, we have observed that in the angiosperm phylogeny \citep{zanne2014three}, the observed clades (\textit{monocotyledoneae}, \textit{magnoliidae}, \textit{superasteridae}, \textit{superrosidae}) have most speciation events occurring on shorter branches (Fig.~\ref{Fig13}), corresponding to higher speciation rates. In contrast, some observed clades in the squamate phylogeny \citep{pyron2013phylogeny}, namely the \textit{iguania} and \textit{anguimorpha} clades, have speciation events occurring on longer branches (Fig.~\ref{Fig12}b,d), indicating a shift from lower to higher speciation rates along the branch lengths on those clades. However, we are aware that this trend on the shift of speciation rates could be affected by sampling bias of extant tips on those reconstructed trees.
	
	In summary, we have demonstrated that our macroevolutionary model with Coxian PH distribution, provides a better fit to empirical phylogenies, when compared to models with other distributions, including exponential and Weibull (Table~\ref{Fig8} and Table~\ref{Fig9}). We conclude that it is necessary to use distributions with sufficient complexity, such as Coxian PH distribution, to provide a better fit to empirical phylogenies.
	
	\begin{acknowledgements}
		We would like to thank the Australian Research Council for funding this research through Discovery Project DP180100352. We also would like to thank Oskar Hagen from ETH 
		Z{\"u}rich for the insight in solving an issue with generating trees using the \textit{TreeSimGM} package.
	\end{acknowledgements}
	
	%
	\section*{Conflict of interest} The authors declare that they have no conflict of interest.
	
	\section{Appendix}
	\label{sec:14}
	\subsection{Equivalent Formula of $q_{n}(i,\beta)$}
	\label{sec:15}
	
	There exists two different formulas of computing the probability of observing $i$ number of left tips given $n$ extant tips on a tree, $q_{n}(i,\beta)$. The first expression includes a product of gamma functions with a normalising constant, $a_{n}(\beta)$, as seen in Eq.~{4} from \cite{aldous1996probability}, while the second expression includes a product of beta functions with a normalising constant, $\hat{a_{n}}(\beta)$, as seen in \textit{maxlik.betasplit} command from \textit{apTreeshape} package (\url{https://github.com/bcm-uga/apTreeshape/blob/master/R/maxlik.betasplit.R}). Here, we show that both expressions are equivalent by showing that both normalising constants are related.
	
	Recall from \cite{aldous1996probability}, we have,
	\begin{equation}
		q_{n}(i,\beta)=\frac{1}{a_{n}(\beta)}\frac{\tau(\beta+i+1)\tau(\beta+n-i+1)}{\tau(i+1)\tau(n-i+1)}, 1\leq i \leq n-1,
		\label{aldousprob}
	\end{equation} 
	where $a_{n}(\beta)$ is a normalizing constant and $\tau(x)$ is the gamma function.
	
	Recall from the \textit{maxlik.betasplit} command, we have,
	\begin{equation}
		\hat{q}_{n}(i,\beta) = \frac{1}{\hat{a_{n}}(\beta)}\frac{B(\beta+i+1,\beta+n-i+1)}{B(i+1,n-i+1)},1\leq i \leq n-1,
		\label{packageprob}
	\end{equation}
	where $\hat{a_{n}}(\beta)$ is a normalizing constant and $B(x,y)$ is beta function.
	
	\begin{proof}
		
		Using the relation between gamma and beta functions where $B(x,y)=\frac{\tau(x)\tau(y)}{\tau(x+y)}$, we can write Eq.~\ref{packageprob} as,
		\begin{align}
			\hat{q}_{n}(i,\beta) &= \frac{1}{\hat{a_{n}(\beta)}} \frac{\frac{\tau(\beta+i+1)\tau(\beta+n-i+1)}{\tau(2\beta+n+2)}}{\frac{\tau(i+1)\tau(n-i+1)}{\tau(n+2)}}\\
			&= \frac{\tau(n+2)}{\hat{a_{n}}(\beta)\tau(2\beta+n+2)}\frac{\tau(\beta+i+1)\tau(\beta+n-i+1)}{\tau(i+1)\tau(n-i+1)}\\
			&= \frac{\tau(n+2)}{\hat{a_{n}}(\beta)\tau(2\beta+n+2)}a_{n}(\beta)q_{n}(i,\beta).
		\end{align}
		
		Hence, $\hat{q}_{n}(i,\beta) = q_{n}(i,\beta)$ if and only if $\frac{1}{\hat{a_{n}}(\beta)}=\frac{\tau(n+2)}{a_{n}(\beta)\tau(2\beta+n+2)}$. That is,
		$\hat{a}_{n}(\beta)=\frac{a_{n}(\beta)\tau(2\beta+n+2)}{\tau(n+2)}$.\qed
	\end{proof}
	
	\subsection{Equivalent Formula of $q_{n}(i,\beta)$ for large $n$ and $i$}
	\label{sec:16}
	
	Here, we show the work to approximate Eq.~\ref{aldousprob} and Eq.~\ref{packageprob} for large $n$ and $i$, where $n$ is the number of extant tips on the tree and $i$ is the number of left tips on the tree. We use this approximation due to computational limitation of evaluating gamma function for large number. From the \textit{maxlik.betasplit} command, we have the following function for large $n$ and $i$,
	
	\begin{eqnarray}
		\hat{q}_{n}(i,\beta) = \frac{1}{\hat{a}_{n}(\beta)}\left(\frac{i}{n}\right)^{\beta}\left(1-\frac{i}{n}\right)^{\beta},
		\label{approx}
	\end{eqnarray}
	where $\hat{a}_{n}(\beta)$ is the normalizing constant for $\hat{q}_{n}(i,\beta)$.
	Here, we show that Eq.~\ref{aldousprob} follows Eq.~\ref{approx} for large $n$ and $i$.
	
	\begin{proof}
		
		Recall the Stirling's approximation for gamma function is given by,
		
		\begin{eqnarray}
			\tau(z) \approx \sqrt{\frac{2\pi}{z}}\left(\frac{z}{e}\right)^{z}.
		\end{eqnarray}
		Then, we claim that,
		
		\begin{eqnarray}
			\frac{\tau(x+\beta+1)}{\tau(x+\alpha+1)} \approx x^{\beta-\alpha} \text{ for large }x.
		\end{eqnarray}
		
		\begin{proof}
			
			By Stirling's approximation with $z=x+\beta$ and $z=x+\alpha$, we have, 
			\begin{align}
				\frac{\tau(x+\beta+1)}{\tau(x+\alpha+1)} &= \frac{(x+\beta)\tau(x+\beta)}{(x+\alpha)\tau(x+\alpha)}\text{, since $x+\beta$ and $x+\alpha \in Z$}\\
				&\approx \frac{(x+\beta)\sqrt{\frac{2\pi}{x+\beta}}\left(\frac{x+\beta}{e}\right)^{x+\beta}}{(x+\alpha)\sqrt{\frac{2\pi}{x+\alpha}}\left(\frac{x+\alpha}{e}\right)^{x+\alpha}}\\
				&= \frac{\sqrt{2\pi(x+\beta)}\left(\frac{x+\beta}{e}\right)^{x+\beta}}{\sqrt{2\pi(x+\alpha)}\left(\frac{x+\alpha}{e}\right)^{x+\alpha}}\\
				&= \left(\frac{x+\beta}{x+\alpha}\right)^{\frac{1}{2}}\frac{(x+\beta)^{x+\beta}}{(x+\alpha)^{x+\alpha}}\frac{1}{e^{\beta-\alpha}}\\
				&= \frac{(x+\beta)^{x+\beta+1/2}}{(x+\alpha)^{x+\alpha+1/2}}\frac{1}{e^{\beta-\alpha}}\\
				&= \frac{(x+\alpha+\beta-\alpha)^{x+\alpha+1/2}}{(x+\alpha)^{x+\alpha+1/2}}\frac{(x+\beta)^{\beta-\alpha}}{e^{\beta-\alpha}}\\
				&= \left(1+\frac{\beta-\alpha}{x+\alpha}\right)^{x+\alpha+1/2}\left(\frac{x+\beta}{x}\right)^{\beta-\alpha}\frac{x^{\beta-\alpha}}{e^{\beta-\alpha}}\\
				&= \left(1+\frac{\beta-\alpha}{x+\alpha}\right)^{x+\alpha+1/2}\left(1+\frac{\beta}{x}\right)^{\beta-\alpha}\left(\frac{x}{e}\right)^{\beta-\alpha}.
			\end{align}
			We observe here that $\left(1+\frac{\beta-\alpha}{x+\alpha}\right)^{x+\alpha+1/2)} \rightarrow e^{\beta-\alpha}$ as $x \rightarrow \infty$ and $\left(1+\frac{\beta}{x}\right)^{\beta-\alpha} \rightarrow 1$ as $x \rightarrow \infty$. Therefore,
			\begin{align}
				\frac{\tau(x+\beta+1)}{\tau(x+\alpha+1)} &\approx e^{\beta-\alpha}\left(\frac{x}{e}			\right)^{\beta-\alpha}\\
				&= x^{\beta-\alpha}.
				\label{add}
			\end{align}\qed
		\end{proof}
		
		Recall that $q_{n}(i,\beta)=\frac{1}{a_{n}(\beta)}\frac{\tau(\beta+i+1)\tau(\beta+n-i+1)}{\tau(i+1)\tau(n-i+1)}$.
		Then, we apply Eq.~\ref{add} for large $n$ and $i$, 
		\begin{align}
			q_{n}(i,\beta) &= \frac{1}{a_{n}(\beta)}\frac{\tau(\beta+i+1)}{\tau(i+1)}\frac{\tau(\beta+n-i+1)}{\tau(n-i+1)}\\
			&\approx \frac{1}{a_{n}(\beta)} i^{\beta}(n-i)^{\beta}\\
			&= \frac{n^{2\beta}}{a_{n}(\beta)}\left(\frac{i}{n}\right)^{\beta}\left(1-\frac{i}{n}\right)^{\beta}.
		\end{align}
		That is, $q_{n}(i,\beta)=\hat{q}_{n}(i,\beta)$ if and only if $\hat{a}_{n}(\beta)=\frac{a_{n}(\beta)}{n^{2\beta}}$.\qed
	\end{proof}
	
\begin{figure}[h!]
		\centering
		\includegraphics[scale=0.5]{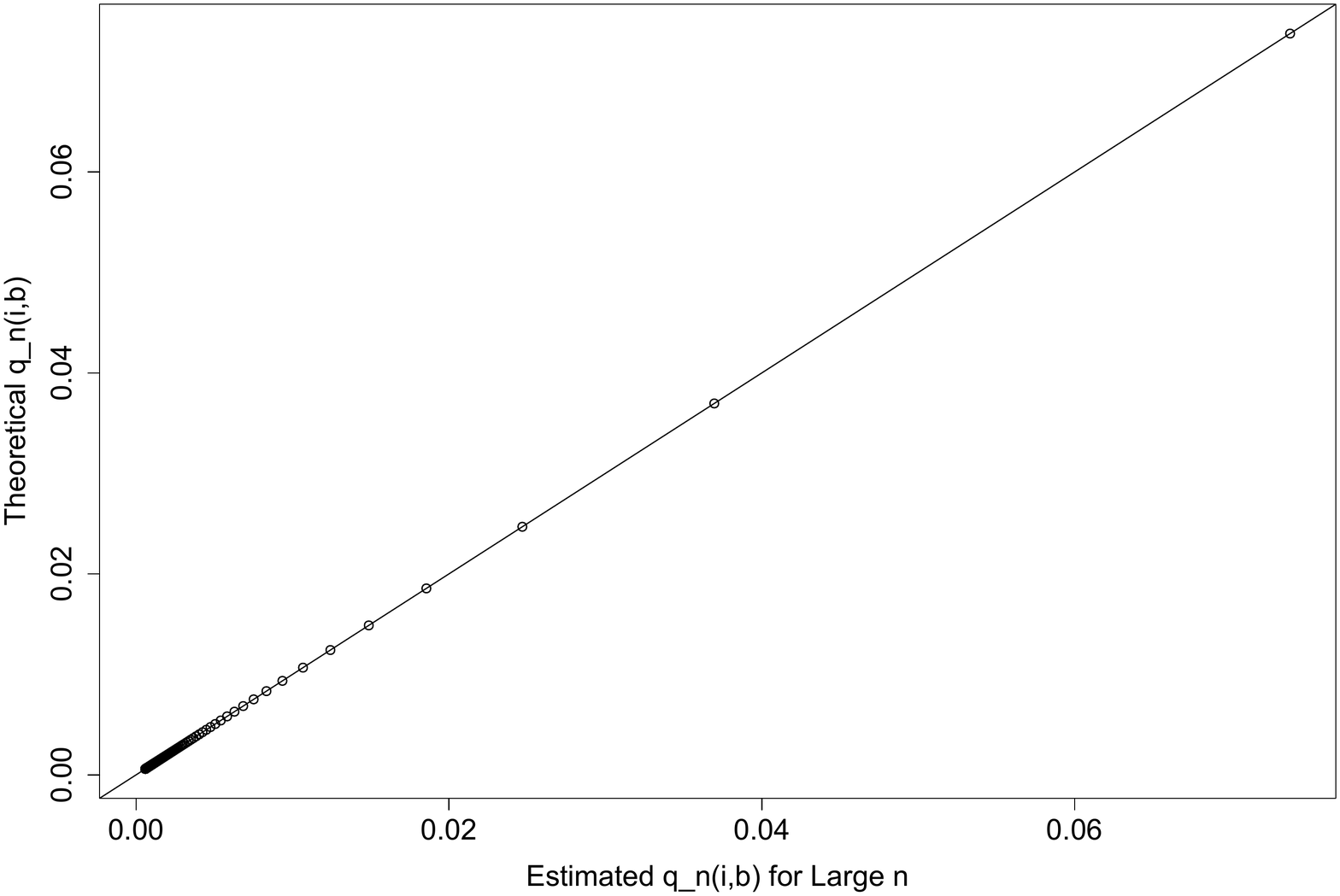}
		\nobreak
		\vspace*{-2mm}
		\caption{Comparison of the probability $q_{n}(i,\beta)$ defined in Eq.~\ref{approx} and Eq.~\ref{aldousprob} for $n = 500$ and $\beta = -1$. The $x$-axis represents the probability defined in Eq.~\ref{approx} while the $y$-axis represents the probability defined in Eq.~\ref{aldousprob}}
		\label{fig2}
	\end{figure}

	To verify the result, we conduct a simulation for $n=500$ and $\beta=-1$ (See Fig.~\ref{fig2}).
	
	\subsection{Expression of First and Second Moments from Coxian PH Distribution}
	\label{sec:17}
	
	In this section, we derive the expressions for first and second moments from a Coxian PH distribution, then we also derive those expressions for the two examples of a Coxian PH distribution used on this paper. The structure of the rate matrix $\mathbf{Q}$ follows the canonical form $3$ described in \cite{okamura2016ph}.
	
	Consider a Coxian PH distribution with four non-absorbing states defined by its rate matrix given as follows,
	
	\begin{eqnarray}
		&\mathbf{Q}=\begin{bmatrix}
			-\lambda_{1} & p_{1}\lambda_{1} & & & \\
			& -\lambda_{2} & p_{2}\lambda_{2} & & \\
			& & -\lambda_{3} & p_{3}\lambda_{3} \\
			& & & & -\lambda_{4}
		\end{bmatrix},
		\label{gencoxian}
	\end{eqnarray}
	
	where $0<p_{1},p_{2},p_{3}\leq 1$.
	Furthermore, we have the condition that $\lambda_1 \geq \lambda_{2} \geq \lambda_{3} \geq \lambda_{4}$ based on the result in \cite{cumani1982canonical} and \cite{dehon1982geometric} for acyclic PH distributions. 
	
	In order to derive the expression of first and second moments of a Coxian PH distribution, we compute the inverse matrix in Eq.~\ref{gencoxian} using the identity matrix of the same size and performing elementary row operations to derive $\left(\mathbf{I}|\mathbf{(Q)^{-1}}\right)$ from $\left(\mathbf{Q}|\mathbf{I}\right)$.
	
	\begin{alignat*}{2}
		\begin{sysmatrix}{rrrr|rrrr}
			-\lambda_{1} & p_{1}\lambda_{1} & 0 & 0 & 1 & 0 & 0 & 0\\
			-0 & -\lambda_{2} & p_{2}\lambda_{2} & 0 & 0 & 1 & 0 & 0 \\
			0 & 0 & -\lambda_{3} & p_{3}\lambda_{3} & 0 & 0 & 1 & 0 \\
			0 & 0 & 0 & -\lambda_{4} & 0 & 0 & 0 & 1
		\end{sysmatrix}
		&\!\begin{aligned}
			&\ro{-\frac{1}{\lambda_{1}}r_1}\\
			&\ro{-\frac{1}{\lambda_{2}}r_2}
		\end{aligned}
		\begin{sysmatrix}{rrrr|rrrr}
			1 & -p_{1} & 0 & 0 & -\frac{1}{\lambda_{1}} & 0 & 0 & 0 \\
			0 & 1 & -p_{2} & 0 & 0 & -\frac{1}{\lambda_{2}} & 0 & 0 \\
			0 & 0 & -\lambda_{3} & p_{3}\lambda_{3} & 0 & 0 & 1 & 0 \\
			0 & 0 & 0 & -\lambda_{4} & 0 & 0 & 0 & 1
		\end{sysmatrix}
		\\
		&\!\begin{aligned}
			&\ro{r_1+p_1r_2}\\
			&\ro{-\frac{1}{\lambda_{3}}r_3}
		\end{aligned}
		\begin{sysmatrix}{rrrr|rrrr}
			1 & 0 & -p_{1}p_{2} & 0 & -\frac{1}{\lambda_{1}} & -\frac{p_{1}}{\lambda_{2}} & 0 & 0 \\
			0 & 1 & -p_{2} & 0 & 0 & -\frac{1}{\lambda_{2}} & 0 & 0\\
			0 & 0 & 1 & -p_{3} & 0 & 0 & -\frac{1}{\lambda_{3}} & 0 \\
			0 & 0 & 0 & -\lambda_{4} & 0 & 0 & 0 &  1
		\end{sysmatrix}
		\\
		&\!\begin{aligned}
			&\ro{r_1+p_1p_2r_3}\\
			&\ro{r_2+p_2r_3}
		\end{aligned}
		\begin{sysmatrix}{rrrr|rrrr}
			1 & 0 & 0 & -p_{1}p_{2}p_{3} & -\frac{1}{\lambda_{1}} & -\frac{p_{1}}{\lambda_{2}} & -\frac{p_{1}p_{2}}{\lambda_{3}} & 0 \\
			0 & 1 & 0 & -p_{2}p_{3} & 0 & -\frac{1}{\lambda_{2}} & -\frac{p_{2}}{\lambda_{3}} & 0 \\
			0 & 0 & 1 & -p_{3} & 0 & 0 & -\frac{1}{\lambda_{3}} & 0 \\
			0 &  0 & 0 & -\lambda_{4} & 0 & 0 & 0 & 1
		\end{sysmatrix}
		\\
		&\!\begin{aligned}
			&\ro{-\frac{1}{\lambda_{4}}r_4}
		\end{aligned}
		\begin{sysmatrix}{rrrr|rrrr}
			1 & 0 & 0 & -p_{1}p_{2}p_{3} & -\frac{1}{\lambda_{1}} & -\frac{p_{1}}{\lambda_{2}} & -\frac{p_{1}p_{2}}{\lambda_{3}} & 0 \\
			0 & 1 & 0 & -p_{2}p_{3} & 0 & -\frac{1}{\lambda_{2}} & -\frac{p_{2}}{\lambda_{3}} & 0 \\
			0 & 0 & 1 & -p_{3} & 0 & 0 & -\frac{1}{\lambda_{3}} & 0 \\
			0 & 0 & 0 & 1 & 0 & 0 & 0 & -\frac{1}{\lambda_{4}}
		\end{sysmatrix}
		\\
		&\!\begin{aligned}
			&\ro{r_1+p_1p_2p_3r_4}\\
			&\ro{r_2+p_2p_3r_4}\\
			&\ro{r_3+p_3r_4}\\
		\end{aligned}
		\begin{sysmatrix}{rrrr|rrrr}
			1 & 0 & 0 & 0 & -\frac{1}{\lambda_{1}} & -\frac{p_{1}}{\lambda_{2}} & -\frac{p_{1}p_{2}}{\lambda_{3}} & -\frac{p_{1}p_{2}p_{3}}{\lambda_{4}} \\
			0 & 1 & 0 & 0 & 0 & -\frac{1}{\lambda_{2}} & -\frac{p_{2}}{\lambda_{3}} & -\frac{p_{2}p_{3}}{\lambda_{4}} \\
			0 &  0 & 1 & 0 & 0 & 0 & -\frac{1}{\lambda_{3}} & -\frac{p_{3}}{\lambda_{4}} \\
			0 & 0 & 0 & 1 & 0 & 0 & 0 & -\frac{1}{\lambda_{4}}
		\end{sysmatrix}.
	\end{alignat*}
	Therefore,
	\begin{align*}
		&\mathbf{Q}^{-1}=\begin{bmatrix}
			-\frac{1}{\lambda_{1}} & -\frac{p_{1}}{\lambda_{2}} & -\frac{p_{1}p_{2}}{\lambda_{3}} & -\frac{p_{1}p_{2}p_{3}}{\lambda_{4}} \\
			0 & -\frac{1}{\lambda_{2}} & -\frac{p_{2}}{\lambda_{3}} & -\frac{p_{2}p_{3}}{\lambda_{4}}\\
			0 & 0 & -\frac{1}{\lambda_{3}} & -\frac{p_{3}}{\lambda_{4}}\\
			0 & 0 & 0 & -\frac{1}{\lambda_{4}}
		\end{bmatrix}
	\end{align*}
	and $\mathbf{Q}\mathbf{Q}^{-1}=\mathbf{I}$ where $\mathbf{I}$ is the identity matrix.
	
	Furthermore,
	\begin{align*}
		\mathbf{Q}^{-2}&=\left(\mathbf{Q}^{-1}\right)^2\\
		&=\begin{bmatrix}
			\frac{1}{\lambda_{1}^2} & \frac{p_{1}}{\lambda_{2}}\left(\frac{1}{\lambda_{1}}+\frac{1}{\lambda_{2}}\right) & \frac{p_{1}p_{2}}{\lambda_{3}}\left(\frac{1}{\lambda_{1}}+\frac{1}{\lambda_{2}}+\frac{1}{\lambda_{3}}\right) & \frac{p_{1}p_{2}p_{3}}{\lambda_{4}}\left(\frac{1}{\lambda_{1}}+\frac{1}{\lambda_{2}}+\frac{1}{\lambda_{3}}+\frac{1}{\lambda_{4}}\right) \\
			0 & \frac{1}{\lambda_{2}^2} & \frac{p_{2}}{\lambda_{3}}\left(\frac{1}{\lambda_{2}}+\frac{1}{\lambda_{3}}\right) & \frac{p_{2}p_{3}}{\lambda_{4}}\left(\frac{1}{\lambda_{2}}+\frac{1}{\lambda_{3}}+\frac{1}{\lambda_{4}}\right)\\
			0 & 0 & \frac{1}{\lambda_{3}^2} & \frac{p_{3}}{\lambda_{4}}\left(\frac{1}{\lambda_{3}}+\frac{1}{\lambda_{4}}\right)\\
			0 & 0 & 0 & \frac{1}{\lambda_{4}^2}
		\end{bmatrix}.
	\end{align*} 
	Hence, the expressions for first and second moments from a Coxian PH distribution with the initial probability distribution $\bm{\alpha}=\left[1, 0, 0, 0 \right]$ and the rate matrix given by Eq.~\ref{gencoxian} are as follows,
	\begin{flalign}
		E_{PH}(X) &= -\bm{\alpha}\mathbf{Q}^{-1}\mathbf{1} \nonumber \\
		&=-\begin{bmatrix}
			1 & 0 & 0 & 0
		\end{bmatrix}
		\begin{bmatrix}
			-\frac{1}{\lambda_{1}} & -\frac{p_{1}}{\lambda_{2}} & -\frac{p_{1}p_{2}}{\lambda_{3}} & -\frac{p_{1}p_{2}p_{3}}{\lambda_{4}} \\
			0 & -\frac{1}{\lambda_{2}} & -\frac{p_{2}}{\lambda_{3}} & -\frac{p_{2}p_{3}}{\lambda_{4}}\\
			0 & 0 & -\frac{1}{\lambda_{3}} & -\frac{p_{3}}{\lambda_{4}}\\
			0 & 0 & 0 & -\frac{1}{\lambda_{4}}
		\end{bmatrix}
		\begin{bmatrix}
			1 \\
			1 \\
			1 \\
			1
		\end{bmatrix}\nonumber \\
		&= \frac{1}{\lambda_{1}}+\frac{p_{1}}{\lambda_{2}}+\frac{p_{1}p_{2}}{\lambda_{3}}+\frac{p_{1}p_{2}p_{3}}{\lambda_{4}}, \nonumber \\ 
		E_{PH}(X)&=\frac{1}{\lambda_{1}}+p_{1}\left(\frac{1}{\lambda_{2}}+p_{2}\left(\frac{1}{\lambda_{3}}+\frac{p_{3}}{\lambda_{4}}\right)\right).
		\label{firstmoment}
	\end{flalign}
	
	\begin{flalign}
		E_{PH}\left(X^{2}\right) &= 2\bm{\alpha}\mathbf{Q}^{-2}\mathbf{1} \nonumber \\
		&=2\begin{bmatrix}
			1 & 0 & 0 & 0
		\end{bmatrix}
		\begin{bmatrix}
			\frac{1}{\lambda_{1}^2} & \frac{p_{1}}{\lambda_{2}}\left(\frac{1}{\lambda_{1}}+\frac{1}{\lambda_{2}}\right) & \frac{p_{1}p_{2}}{\lambda_{3}}\left(\frac{1}{\lambda_{1}}+\frac{1}{\lambda_{2}}+\frac{1}{\lambda_{3}}\right) & \frac{p_{1}p_{2}p_{3}}{\lambda_{4}}\left(\frac{1}{\lambda_{1}}+\frac{1}{\lambda_{2}}+\frac{1}{\lambda_{3}}+\frac{1}{\lambda_{4}}\right) \\
			0 & \frac{1}{\lambda_{2}^2} & \frac{p_{2}}{\lambda_{3}}\left(\frac{1}{\lambda_{2}}+\frac{1}{\lambda_{3}}\right) & \frac{p_{2}p_{3}}{\lambda_{4}}\left(\frac{1}{\lambda_{2}}+\frac{1}{\lambda_{3}}+\frac{1}{\lambda_{4}}\right)\\
			0 & 0 & \frac{1}{\lambda_{3}^2} & \frac{p_{3}}{\lambda_{4}}\left(\frac{1}{\lambda_{3}}+\frac{1}{\lambda_{4}}\right)\\
			0 & 0 & 0 & \frac{1}{\lambda_{4}^2}
		\end{bmatrix}
		\begin{bmatrix}
			1 \\
			1 \\
			1 \\
			1
		\end{bmatrix}, \nonumber \\
		E_{PH}\left(X^{2}\right)&= 2\left[\frac{1}{\lambda_{1}^{2}}+\frac{p_{1}}{\lambda_{2}}\left(\frac{1}{\lambda_{1}}+\frac{1}{\lambda_{2}}\right)+\frac{p_{1}p_{2}}{\lambda_{3}}\left(\frac{1}{\lambda_{1}}+\frac{1}{\lambda_{2}}+\frac{1}{\lambda_{3}}\right)+\frac{p_{1}p_{2}p_{3}}{\lambda_{4}}\left(\frac{1}{\lambda_{1}}+\frac{1}{\lambda_{2}}+\frac{1}{\lambda_{3}}+\frac{1}{\lambda_{4}}\right)\right].
		\label{secondmoment}
	\end{flalign}
	
	\subsection{Expression of First and Second Moments of Our Model Corresponding to Weibull Distribution with Shape Parameter Smaller Than One}
	\label{sec:18}
	
	We define the following sub-rate matrix $\mathbf{Q}$ which follows the structure of the canonical form $3$ for acyclic PH distributions from \cite{okamura2016ph} and corresponds to Eq.~\ref{gencoxian} with the following parameter changes. We find that this matrix structure corresponds to a Weibull distribution for shape parameter $\Phi < 1$,
	\begin{align}
		&\lambda_{1}=z,\lambda_{2}=1+x,\lambda_{3}=1+x^{2},\lambda_{4}=x^{3} \nonumber \\
		&p_{1}=1-y,p_{2}=1-y^{2},p_{3}=1-y^{3}
		\label{changepar}
	\end{align}
	where $x,y,\text{and } z$ are parameters from the following rate matrix,
	\begin{align}
		\mathbf{Q}=
		\begin{bmatrix}
			-z & (1-y)z & 0 & 0\\
			0 & -(1+x) & \left(1-y^2\right)(1+x) & 0\\
			0 & 0 & -\left(1+x^2\right) & \left(1-y^3\right)\left(1+x^2\right)\\
			0 & 0 & 0 & -x^3
		\end{bmatrix},
		\label{ourmodel}
	\end{align}
	where $0<x\leq 1$, $0<y<1$ and $z\geq 2$.
	
	Using Eq.~\ref{firstmoment}, Eq.~\ref{secondmoment}, and Eq.~\ref{changepar}, we get the expressions for the first and second moments for our PH distribution with the rate matrix shown in Eq.~\ref{ourmodel} given by, 
	\begin{eqnarray}
		E_{PH}(X) &=&\frac{1}{z}+(1-y)\left(\frac{1}{1+x}+\left(1-y^2\right)\left(\frac{1}{1+x^2}+\frac{1-y^3}{x^3}\right)\right),
		\label{MeanZ}
	\end{eqnarray}
	
	\begin{eqnarray}
		E_{PH}\left(X^{2}\right)&=&2\left[\frac{1}{z^{2}}+\frac{1-y}{1+x}\left(\frac{1}{z}+\frac{1}{1+x}\right)+\frac{(1-y)\left(1-y^2\right)}{1+x^2}\left(\frac{1}{z}+\frac{1}{1+x}+\frac{1}{1+x^2}\right)\right.\nonumber \\
		&&\left. +\frac{(1-y)\left(1-y^2\right)\left(1-y^3\right)}{x^3}\left(\frac{1}{z}+\frac{1}{1+x}+\frac{1}{1+x^2}+\frac{1}{x^3}\right)\right].
		\label{secmomentz}
	\end{eqnarray}
	
	Later, we find $x,y,\text{ and }z$ values that corresponds to same first and second moment values from Weibull distribution under different shape ($\Phi<1$) and scale ($\psi$) parameters, induced as speciation and extinction rates separately as shown by Figure $2$ on \cite{Hagen2015}. From observation, it suffices to assume $z=2$ in order to match the first and second moments from the corresponding Weibull distributions. From Eq.~\ref{MeanZ} and Eq.~\ref{secmomentz}, we have the following expressions for first and second moments for $z=2$,
	\begin{eqnarray}
		E_{PH}(X) &=&\frac{1}{2}+(1-y)\left(\frac{1}{1+x}+\left(1-y^2\right)\left(\frac{1}{1+x^2}+\frac{1-y^3}{x^3}\right)\right),
		\label{MeanZ2}
	\end{eqnarray}
	
	\begin{eqnarray}
		E_{PH}\left(X^{2}\right)&=&\frac{1}{2}+\frac{2(1-y)}{1+x}\left(\frac{1}{2}+\frac{1}{1+x}\right)+\frac{2(1-y)\left(1-y^2\right)}{1+x^2}\left(\frac{1}{2}+\frac{1}{1+x}+\frac{1}{1+x^2}\right)\nonumber \\
		&& +\frac{2(1-y)\left(1-y^2\right)\left(1-y^3\right)}{x^3}\left(\frac{1}{2}+\frac{1}{1+x}+\frac{1}{1+x^2}+\frac{1}{x^3}\right).
		\label{secmomentz2}
	\end{eqnarray}
	
	\subsection{Expression of First and Second Moments of Our Model Corresponding to Weibull Distribution with Shape Parameter Larger Than or Equal One}
	\label{sec:19}
	
	We define the following rate matrix $\mathbf{Q}$ which follows the structure of the canonical form $3$ for acyclic PH distributions from \cite{okamura2016ph} and corresponds to Eq.~\ref{gencoxian} with the following parameter changes. We find that this matrix structure corresponds with Weibull distribution for shape parameter $\Phi \geq 1$
	\begin{align}
		&\lambda_{1}=1+x^{3},\lambda_{2}=1+x^{2},\lambda_{3}=1+x,\lambda_{4}=z \nonumber \\
		&p_{1}=1-y^{4},p_{2}=1-y^{3},p_{3}=1-y^{2}
		\label{changepar1}
	\end{align}
	where $x,y,\text{and } z$ are parameters from the following rate matrix,
	\begin{align}
		\mathbf{Q}=
		\begin{bmatrix}
			-\left(1+x^{3}\right) & \left(1-y^{4}\right)\left(1+x^{3}\right) & 0 & 0\\
			0 & -\left(1+x^{2}\right) &  \left(1-y^{3}\right)\left(1+x^{2}\right) & 0\\
			0 & 0 & -(1+x) & \left(1-y^2\right)(1+x)\\
			0 & 0 & 0 & -z
		\end{bmatrix},
		\label{ourmodel1}
	\end{align}
	where $0<x\leq 1$, $0<y<1$ and $z\geq 2$.
	
	Using Eq.~\ref{firstmoment}, Eq.~\ref{secondmoment}, and Eq.~\ref{changepar}, we get the expressions for the first and second moments for our PH distribution with the rate matrix shown in Eq.~\ref{ourmodel1} given by, 
	\begin{eqnarray}
		E_{PH}(X) &=&\frac{1}{1+x^{3}}+\left(1-y^{4}\right)\left(\frac{1}{1+x^{2}}+\left(1-y^3\right)\left(\frac{1}{1+x}+\frac{1-y^2}{z}\right)\right),
		\label{MeanZ1}
	\end{eqnarray}
	
	\begin{eqnarray}
		E_{PH}\left(X^{2}\right)&=&\frac{2}{\left(1+x^3\right)^2}+\frac{2\left(1-y^4\right)}{1+x^2}\left(\frac{1}{1+x^3}+\frac{1}{1+x^2}\right)+\frac{2\left(1-y^4\right)\left(1-y^3\right)}{1+x}\nonumber \\
		&&\left(\frac{1}{1+x^3}+\frac{1}{1+x^2}+\frac{1}{1+x}\right) +\frac{2\left(1-y^4\right)\left(1-y^3\right)\left(1-y^2\right)}{z}\nonumber\\
		&&\left(\frac{1}{1+x^3}+\frac{1}{1+x^2}+\frac{1}{1+x}+\frac{1}{z}\right).
		\label{secmomentz1}
	\end{eqnarray}
	
	Later, we find $x,y,\text{ and }z$ values that correspond to the same first and second moment values from Weibull distribution under different shape ($\Phi\geq 1$) and scale ($\psi$) parameters for speciation and extinction rates separately, as shown by Figure $2$ on \cite{Hagen2015}. It suffices to assume $z=2$ in order to match the first and second moments from the corresponding Weibull distributions with shape parameter $\Phi>1$. By Eq.~\ref{MeanZ1} and Eq.~\ref{secmomentz1}, we have the following expressions for first and second moments for $z=2$,
	\begin{eqnarray}
		E_{PH}(X) &=&\frac{1}{1+x^{3}}+\left(1-y^{4}\right)\left(\frac{1}{1+x^{2}}+\left(1-y^3\right)\left(\frac{1}{1+x}+\frac{1-y^2}{2}\right)\right),
		\label{MeanZ3}
	\end{eqnarray}
	
	\begin{eqnarray}
		E_{PH}\left(X^{2}\right)&=&\frac{2}{\left(1+x^3\right)^2}+\frac{2\left(1-y^4\right)}{1+x^2}\left(\frac{1}{1+x^3}+\frac{1}{1+x^2}\right)+\frac{2\left(1-y^4\right)\left(1-y^3\right)}{1+x}\nonumber \\
		&&\left(\frac{1}{1+x^3}+\frac{1}{1+x^2}+\frac{1}{1+x}\right) +\left(1-y^4\right)\left(1-y^3\right)\left(1-y^2\right)\nonumber\\
		&&\left(\frac{1}{1+x^3}+\frac{1}{1+x^2}+\frac{1}{1+x}+\frac{1}{2}\right).
		\label{secmomentz3}
	\end{eqnarray}
	\newpage
	\begin{figure}[!htbp]
		\centering
		\begin{subfigure}[!htbp]{0.55\textwidth}
			\centering
			\includegraphics[width=1\linewidth]{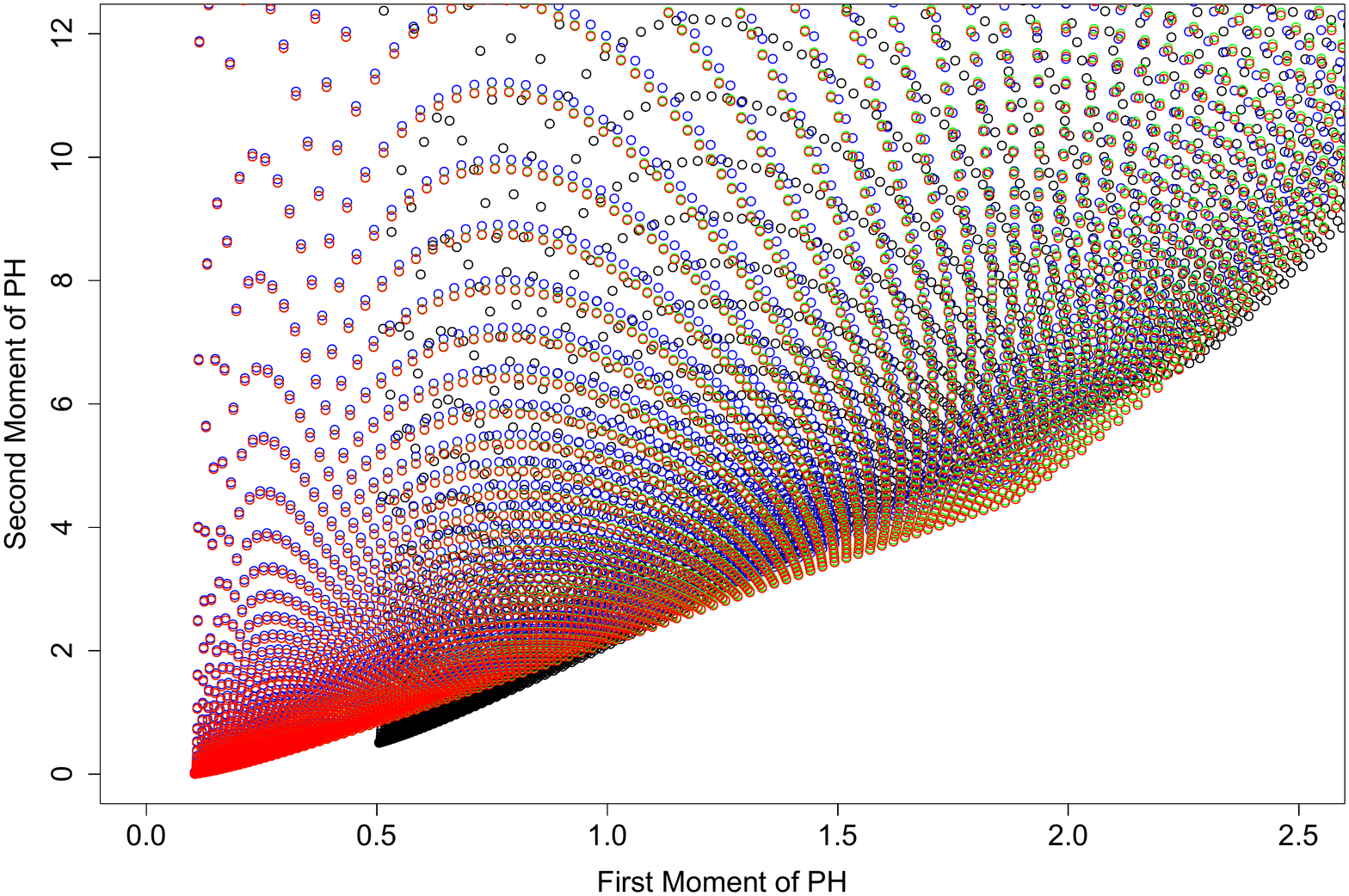}
		\end{subfigure}%
		~ 
		\begin{subfigure}[!htbp]{0.55\textwidth}
			\centering
			\includegraphics[width=1\linewidth]{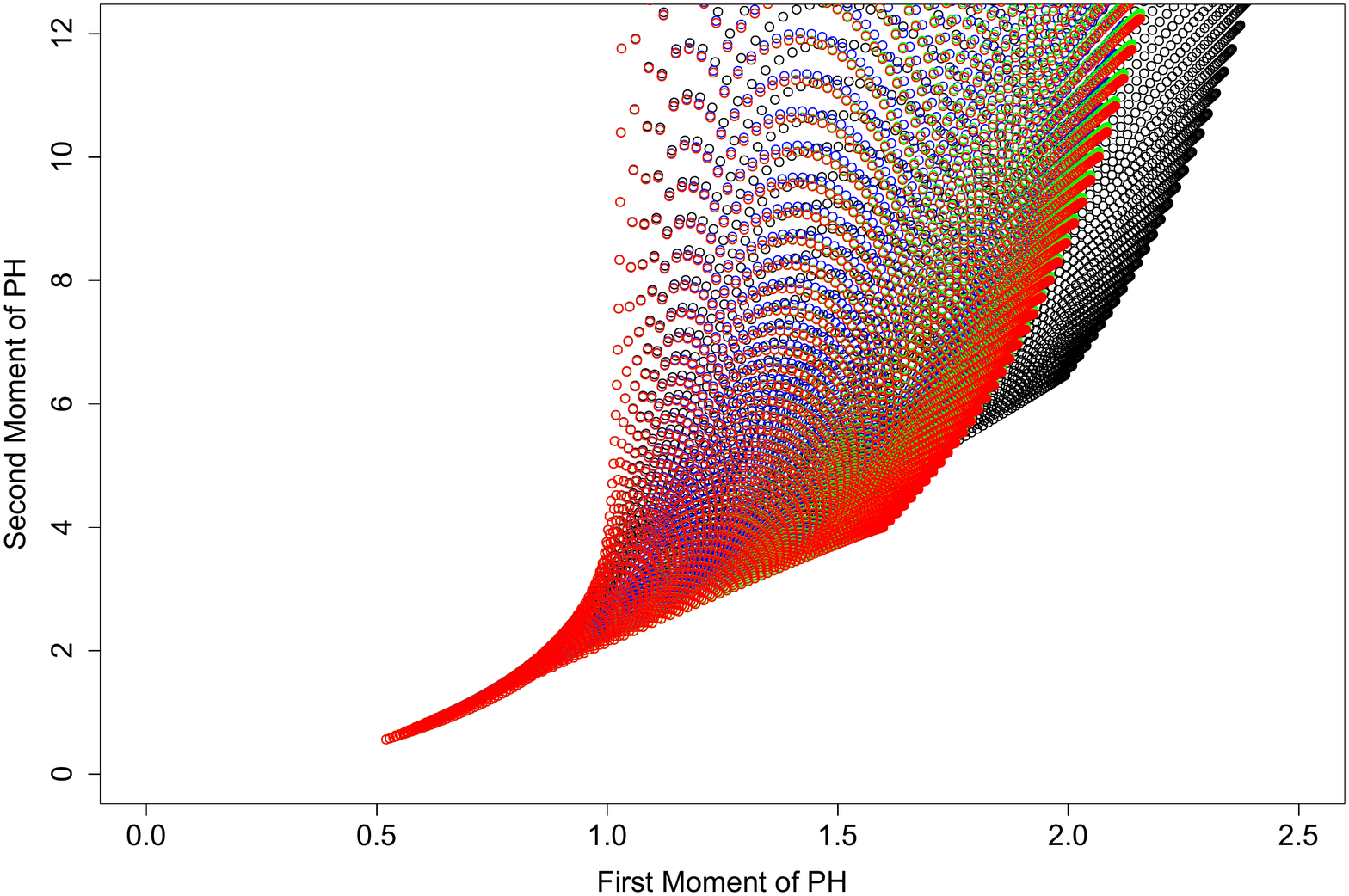}
		\end{subfigure}
		\caption{The left figure area of the first and second moments of Eq.~\ref{MeanZ} and Eq.~\ref{secmomentz} for various $z$ values while the right figure shows the first and second moments of Eq.~\ref{MeanZ1} and Eq.~\ref{secmomentz1}}
		\label{Fig13alt}
	\end{figure}
	
	\subsection{Speciation Rate Following Weibull Distribution}
	\label{sec:20}
	
	As observed in Figure $2$ on \cite{Hagen2015}, in order to test that speciation rate affects tree shape, they have used Weibull distribution with mean time to speciation events $\mu_{s}$ equals to $1$ and shape parameters $\Phi_{s} \in \{0.2,0.5,1,1.3\}$ (Note: We have substituted $\Phi_{s}=0.1$ with $\Phi_{s}=0.2$ since the former produces a very large variance but both values still imply a decreasing speciation rate). The mean time to speciation events $\mu_{s}$ and shape parameters $\Phi_{s}$ correspond to the following scale parameters $\psi_{s} \in \{0.0083,0.5,1,1.08\}$. Their first and second moment values are given by the following set, $\left(E(X),E\left(X^2\right)\right)=\{(1,252),(1,6),(1,2),(1,1.6)\}$. 
	
	Matching each pair of $\left(E(X),E\left(X^2\right)\right)$ with Eq.~\ref{MeanZ2} and Eq.~\ref{secmomentz2} for the case $\Phi_{s}<1$ and Eq.~\ref{MeanZ3} and Eq.~\ref{secmomentz3} for the case $\Phi_{s}\geq 1$,
	
	\noindent we derive $(x,y) \in \{(0.32,0.67),(0.52,0.79),(0.61,0.99),(1,0.81)\}$.
	
	\subsection{Extinction Rate Following Weibull Distribution}
	\label{sec:21}
	
	Similarly, as seen on Figure $2$ on \cite{Hagen2015}, In order to test that extinction rate affects tree balance, they have used Weibull distribution with mean time to extinction events $\mu_{e}$ equals to $2$ and shape parameters $\Phi_{e} \in \{0.7,1,1.3,1.5\}$. The mean time to extinction events $\mu_{e}$ and shape parameters $\Phi_{e}$ correspond to the following scale parameters $\psi_{e} \in \{1.58,2,2.17,2.22\}$. Their first and second moment values are given by the following set, $\left(E(X),E\left(X^2\right)\right)=\{(2,12.55),(2,8),(2,6.43),(2,5.87)\}$. 
	
	Matching each pair of $\left(E(X),E\left(X^2\right)\right)$ with Eq.~\ref{MeanZ2} and Eq.~\ref{secmomentz2} for the case $\Phi_{s}<1$ and Eq.~\ref{MeanZ3} and Eq.~\ref{secmomentz3} for the case $\Phi_{s}\geq 1$, 
	
	\noindent we derive $(x,y) \in \{(0.64,0.55),(0.86,0.43),(1, 0.14),(1,0.36)\}$.
	
	\subsection{Fitting Branch Length Data to Our Model with Five Non-absorbing States}
	\label{sec:22}
	
	Here, we show that adding more non-absorbing states in our model from Eq.~\ref{ourmodel} does not significantly improve how fit the model to the distribution of simulated branch length data. 
	
	\begin{figure}[!htbp]
		\centering     
		\captionsetup{width=\linewidth}
		\includegraphics[scale = 0.4]{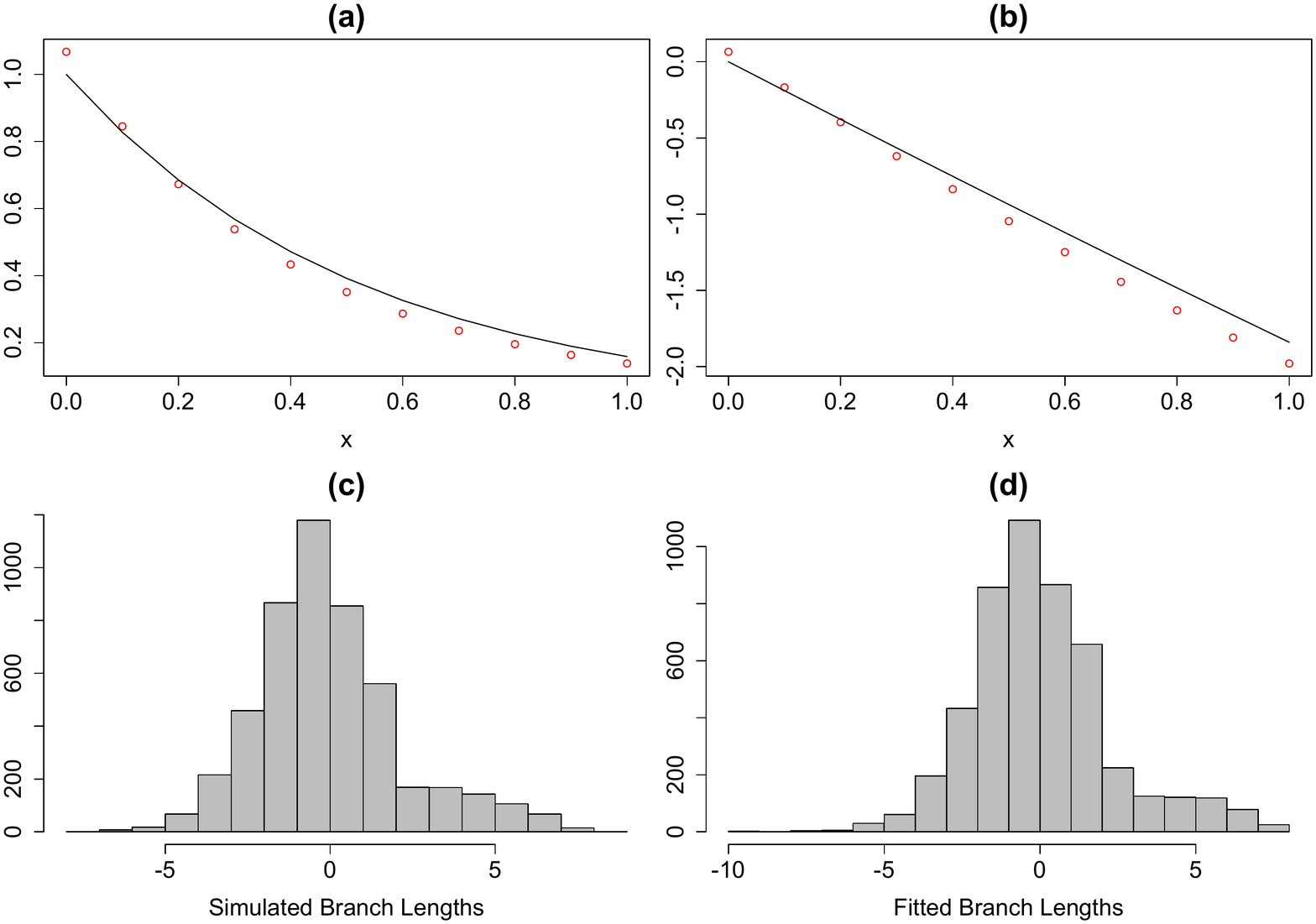}
		\caption{The comparison of density functions between our model as defined in  Eq.~\ref{ourmodel} using fitted parameter values with five non-absorbing states and a simulated four-state general Coxian PH distribution used to generate the branch lengths (Fig.~\ref{fit5states}a,b). The black line represents the density function (Fig.~\ref{fit5states}a) and the log density function (Fig.~\ref{fit5states}b) of the latter distribution, while the red dots represent the density function of the fitted PH distribution. The histograms of the simulated and fitted branch length distributions, shown in log scale, are displayed in Figure~\ref{fit5states}c,d respectively}
		\label{fit5states}
	\end{figure}

	\bibliographystyle{spbasic}      
	\bibliography{References}   

\begin{thebibliography}{46}
\providecommand{\natexlab}[1]{#1}
\providecommand{\url}[1]{{#1}}
\providecommand{\urlprefix}{URL }
\expandafter\ifx\csname urlstyle\endcsname\relax
  \providecommand{\doi}[1]{DOI~\discretionary{}{}{}#1}\else
  \providecommand{\doi}{DOI~\discretionary{}{}{}\begingroup
  \urlstyle{rm}\Url}\fi
\providecommand{\eprint}[2][]{\url{#2}}

\bibitem[{Akaike(1998)}]{akaike1998information}
Akaike H (1998) Information theory and an extension of the maximum likelihood
  principle. In: Selected papers of {H}irotugu {A}kaike, Springer, pp 199--213

\bibitem[{Aldous(1996)}]{aldous1996probability}
Aldous DJ (1996) Probability distributions on cladograms. In: Random discrete
  structures, Springer, pp 1--18

\bibitem[{Aldous(2001)}]{aldous2001}
Aldous DJ (2001) Stochastic models and descriptive statistics for phylogenetic
  trees, from {Y}ule to today. Stat Sci 16(1):23--34

\bibitem[{Anderson and Burnham(2004)}]{anderson2004model}
Anderson D, Burnham K (2004) Model selection and multi-model inference. Second
  NY: Springer-Verlag 63(2020):10

\bibitem[{Asmussen et~al.(1996)Asmussen, Nerman, and Olsson}]{asmussen1996}
Asmussen S, Nerman O, Olsson M (1996) Fitting phase-type distributions via the
  {E}{M} algorithm. Scand J Stat 23:419--441

\bibitem[{Bellman and Harris(1948)}]{bellman1948theory}
Bellman R, Harris TE (1948) On the theory of age-dependent stochastic branching
  processes. Proceedings of the National Academy of Sciences of the United
  States of America 34(12):601

\bibitem[{Bortolussi et~al.(2006)Bortolussi, Durand, Blum, and
  Fran{\c{c}}ois}]{bortolussi2006aptreeshape}
Bortolussi N, Durand E, Blum M, Fran{\c{c}}ois O (2006) ap{T}reeshape:
  statistical analysis of phylogenetic tree shape. Bioinformatics
  22(3):363--364

\bibitem[{Byrd et~al.(1995)Byrd, Lu, Nocedal, and Zhu}]{byrd1995limited}
Byrd RH, Lu P, Nocedal J, Zhu C (1995) A limited memory algorithm for bound
  constrained optimization. SIAM J Sci Comput 16(5):1190--1208

\bibitem[{Cumani(1982)}]{cumani1982canonical}
Cumani A (1982) On the canonical representation of homogeneous {M}arkov
  processes modelling failure-time distributions. Microelectron Reliab
  22(3):583--602

\bibitem[{Dehon and Latouche(1982)}]{dehon1982geometric}
Dehon M, Latouche G (1982) A geometric interpretation of the relations between
  the exponential and generalized {E}rlang distributions. Adv Appl Probab
  14(4):885--897

\bibitem[{Etienne et~al.(2012)Etienne, Haegeman, Stadler, Aze, Pearson, Purvis,
  and Phillimore}]{etienne2012diversity}
Etienne RS, Haegeman B, Stadler T, Aze T, Pearson PN, Purvis A, Phillimore AB
  (2012) Diversity-dependence brings molecular phylogenies closer to agreement
  with the fossil record. Proceedings of the Royal Society B: Biological
  Sciences 279(1732):1300--1309

\bibitem[{FitzJohn(2012)}]{fitzjohn2012diversitree}
FitzJohn RG (2012) Diversitree: comparative phylogenetic analyses of
  diversification in {R}. Methods Ecol Evol 3(6):1084--1092

\bibitem[{Hagen and Stadler(2018)}]{hagen2018treesimgm}
Hagen O, Stadler T (2018) Tree{S}im{G}{M}: simulating phylogenetic trees under
  general {B}ellman--{H}arris models with lineage-specific shifts of speciation
  and extinction in {R}. Methods Ecol Evol 9(3):754--760

\bibitem[{Hagen et~al.(2015)Hagen, Hartmann, Steel, and Stadler}]{Hagen2015}
Hagen O, Hartmann K, Steel M, Stadler T (2015) {Age-dependent speciation can
  explain the shape of empirical phylogenies}. Syst Biol 64(3):432--440,
  \doi{10.1093/sysbio/syv001},
  \urlprefix\url{https://doi.org/10.1093/sysbio/syv001},
  \eprint{http://oup.prod.sis.lan/sysbio/article-pdf/64/3/432/24587693/syv001.pdf}

\bibitem[{Harvey and Pagel(1991)}]{harvey1991comparative}
Harvey PH, Pagel MD (1991) The comparative method in evolutionary biology, vol
  239. Oxford University Press

\bibitem[{Huson and Scornavacca(2012)}]{huson2012dendroscope}
Huson DH, Scornavacca C (2012) Dendroscope 3: an interactive tool for rooted
  phylogenetic trees and networks. Syst Biol 61(6):1061--1067

\bibitem[{Lambert and Stadler(2013)}]{lambert2013birth}
Lambert A, Stadler T (2013) Birth--death models and coalescent point processes:
  the shape and probability of reconstructed phylogenies. Theor Popul Biol
  90:113--128

\bibitem[{Louca and Pennell(2020)}]{louca2020extant}
Louca S, Pennell MW (2020) Extant timetrees are consistent with a myriad of
  diversification histories. Nature 580(7804):502--505

\bibitem[{Maddison et~al.(2007)Maddison, Midford, and
  Otto}]{maddison2007estimating}
Maddison WP, Midford PE, Otto SP (2007) Estimating a binary character's effect
  on speciation and extinction. Syst Biol 56(5):701--710

\bibitem[{Marshall and McClean(2004)}]{marshall2004using}
Marshall AH, McClean SI (2004) Using {C}oxian phase-type distributions to
  identify patient characteristics for duration of stay in hospital. Health
  Care Manag Sci 7(4):285--289

\bibitem[{Morlon(2014)}]{morlon2014phylogenetic}
Morlon H (2014) Phylogenetic approaches for studying diversification. Ecol Lett
  17(4):508--525

\bibitem[{Morlon et~al.(2010)Morlon, Potts, and Plotkin}]{morlon2010inferring}
Morlon H, Potts MD, Plotkin JB (2010) Inferring the dynamics of
  diversification: a coalescent approach. PLoS biology 8(9):e1000493

\bibitem[{Morlon et~al.(2011)Morlon, Parsons, and
  Plotkin}]{morlon2011reconciling}
Morlon H, Parsons TL, Plotkin JB (2011) Reconciling molecular phylogenies with
  the fossil record. Proc Natl Acad Sci USA 108(39):16327--16332

\bibitem[{Nee et~al.(1992)Nee, Mooers, and Harvey}]{nee1992tempo}
Nee S, Mooers AO, Harvey PH (1992) Tempo and mode of evolution revealed from
  molecular phylogenies. Proc Natl Acad Sci USA 89(17):8322--8326

\bibitem[{Nee et~al.(1994{\natexlab{a}})Nee, Holmes, May, and
  Harvey}]{nee1994extinction}
Nee S, Holmes EC, May RM, Harvey PH (1994{\natexlab{a}}) Extinction rates can
  be estimated from molecular phylogenies. Phil Trans R Soc Lond B
  344(1307):77--82

\bibitem[{Nee et~al.(1994{\natexlab{b}})Nee, May, and
  Harvey}]{nee1994reconstructed}
Nee S, May RM, Harvey PH (1994{\natexlab{b}}) The reconstructed evolutionary
  process. Phil Trans R Soc Lond B 344(1309):305--311

\bibitem[{Nelder and Mead(1965)}]{nelder1965simplex}
Nelder JA, Mead R (1965) A simplex method for function minimization. Comput J
  7(4):308--313

\bibitem[{Neuts(1975)}]{neuts1975probability}
Neuts MF (1975) Probability distributions of phase-type. Liber Amicorum Prof
  Emeritus H Florin, Department of Mathematics, University of Louvain

\bibitem[{Neuts(1981)}]{neuts1981}
Neuts MF (1981) Matrix-geometric solutions in stochastic models: an algorithmic
  approach. Johns Hopkins University Press Baltimore

\bibitem[{Okamura and Dohi(2016)}]{okamura2016ph}
Okamura H, Dohi T (2016) Ph fitting algorithm and its application to
  reliability engineering. J Oper Res Soc Japan 59(1):72--109

\bibitem[{Paradis et~al.(2004)Paradis, Claude, and Strimmer}]{paradis2004ape}
Paradis E, Claude J, Strimmer K (2004) A{P}{E}: analyses of phylogenetics and
  evolution in {R} language. Bioinformatics 20(2):289--290

\bibitem[{Pawitan(2001)}]{pawitan2001all}
Pawitan Y (2001) In all likelihood: statistical modelling and inference using
  likelihood. Oxford University Press

\bibitem[{Phillimore and Price(2008)}]{phillimore2008density}
Phillimore AB, Price TD (2008) Density-dependent cladogenesis in birds. PLoS
  biology 6(3):e71

\bibitem[{Pybus and Harvey(2000)}]{pybus2000}
Pybus OG, Harvey PH (2000) Testing macro--evolutionary models using incomplete
  molecular phylogenies. Proc Royal Soc B 267(1459):2267--2272

\bibitem[{Pyron et~al.(2013)Pyron, Burbrink, and Wiens}]{pyron2013phylogeny}
Pyron RA, Burbrink FT, Wiens JJ (2013) A phylogeny and revised classification
  of squamata, including 4161 species of lizards and snakes. BMC Evol Biol
  13(1):93

\bibitem[{Quental and Marshall(2010)}]{quental2010diversity}
Quental TB, Marshall CR (2010) Diversity dynamics: molecular phylogenies need
  the fossil record. Trends Ecol Evol 25(8):434--441

\bibitem[{Rabosky(2006)}]{rabosky2006likelihood}
Rabosky DL (2006) Likelihood methods for detecting temporal shifts in
  diversification rates. Evolution 60(6):1152--1164

\bibitem[{Rabosky and Lovette(2008)}]{rabosky2008density}
Rabosky DL, Lovette IJ (2008) Density-dependent diversification in north
  american wood warblers. Proceedings of the Royal Society B: Biological
  Sciences 275(1649):2363--2371

\bibitem[{Ricklefs(2007)}]{ricklefs2007estimating}
Ricklefs RE (2007) Estimating diversification rates from phylogenetic
  information. Trends Ecol Evol 22(11):601--610

\bibitem[{Simpson(1944)}]{simpson1944tempo}
Simpson GG (1944) Tempo and mode in evolution. Columbia University Press

\bibitem[{Stadler(2013)}]{stadler2013recovering}
Stadler T (2013) Recovering speciation and extinction dynamics based on
  phylogenies. J Evol Biol 26(6):1203--1219

\bibitem[{Stanley(1998)}]{stanley1998macroevolution}
Stanley SM (1998) Macroevolution: pattern and process. Johns Hopkins University
  Press

\bibitem[{Thummler et~al.(2006)Thummler, Buchholz, and
  Telek}]{thummler2006novel}
Thummler A, Buchholz P, Telek M (2006) A novel approach for phase-type fitting
  with the {E}{M} algorithm. IEEE Trans Dependable Sec Comput 3(3):245--258

\bibitem[{Verbelen(2013)}]{verbelen2013}
Verbelen R (2013) Phase-type distributions \& mixtures of erlangs. PhD thesis,
  University of Leuven

\bibitem[{Yule(1925)}]{yule1925ii}
Yule GU (1925) Ii.—{A} mathematical theory of evolution, based on the
  conclusions of dr. jc willis, fr s. Phil Trans R Soc Lond B
  213(402-410):21--87

\bibitem[{Zanne et~al.(2014)Zanne, Tank, Cornwell, Eastman, Smith, FitzJohn,
  McGlinn, O’Meara, Moles, Reich et~al.}]{zanne2014three}
Zanne AE, Tank DC, Cornwell WK, Eastman JM, Smith SA, FitzJohn RG, McGlinn DJ,
  O’Meara BC, Moles AT, Reich PB, et~al. (2014) Three keys to the radiation
  of angiosperms into freezing environments. Nature 506(7486):89--92

\end{thebibliography}
	
	
\end{document}